\documentclass[pra, twocolumn, amsmath,amssymb, nofootinbib, floatfix ]{revtex4-1}

\usepackage{multirow}
\usepackage{graphicx}
\usepackage{amsfonts,tikz}  
\usetikzlibrary{matrix}
\usepackage{epstopdf}
\usepackage[version=3]{mhchem}
\usepackage{color}
\usepackage{hyperref}
\usepackage{enumitem}
\usepackage[scr=boondoxo,scrscaled=1.25]{mathalfa}
\usepackage{scalerel}
\usepackage{mathbbol}

\hypersetup{%
  pdfpagemode=None, 
  pdfstartpage=1,
  pdfstartview=FitH,
  pdfmenubar=true,
  pdftoolbar=true,
  colorlinks = true,
  linkcolor=blue,
  citecolor=blue,
  urlcolor = blue,
  bookmarksopen=false
}

\newcommand{\Op}[1]{\ensuremath{\boldsymbol{\mathsf{\hat{#1}}}}}

\newcommand{\Bra}[1]{\ensuremath{\left\langle #1 \right\vert}}
\newcommand{\KetBra}[2]{\Ket{#1}\kern-0.1em\Bra{#2}}
\newcommand{\Ket}[1]{\ensuremath{\left\vert #1 \right\rangle}}

\definecolor{Pantone268}{cmyk}{0.82,1.0,0.0,0.12}
\definecolor{HunterOrange}{cmyk}{0.0,0.55,1.0,0.0}

\begin{document}


\title{Quantum control of entangled photon-pair generation in electron-atom\- 
collisions driven by laser-synthesized free-electron wave packets}

\date{\today}

\author{R.~Esteban Goetz}
\altaffiliation{$\!\!$esteban.goetz@drake.edu}
\affiliation{Department of  Physics and Astronomy, Drake University, Des Moines, IA 50311, USA}

\author{Klaus Bartschat}
\affiliation{Department of  Physics and Astronomy, Drake University, Des Moines, IA 50311, USA}

\begin{abstract}
We propose an extension of coherent control using laser-synthesized free-electron matter waves.
In contrast to coherent control schemes exploiting optical coherences to steer the dynamics of matter waves, we analyze the opposite 
and investigate the control of quantum light emission driven by laser-sculpted coherent free-electron matter waves.  
We apply this concept to the control of entangled photon-pair emission in electron-atom collisions, in which   
the incident electron wave packet, colliding with a target atom $B$, is engineered by interferometric resonantly-enhanced 
multi\-photon ionization of a parent atom $A$. Each ionization pathway leads to electron wave packets that 
coherently interfere during temporal evolution in the continuum. Their mutual
coherence can be controlled by adjusting the relative phases
or time delays of the frequency components of the ionizing field contributing to the interfering pathways. 
We report coherent control of entangled photon-pair generation 
in radiative photo-cascade emission upon decay of the target atom after inelastic excitation triggered by the 
collision with the synthesized electron wave packet.                      
\end{abstract}

\maketitle

Coherence~\cite{TannorJCP85} and quantum 
interferences~\cite{shapiro2012quantum,BrumerReportsPhys} are the cornerstones 
of coherent control of photo-induced processes. A prototypical example is   
two-pathway coherent control of photoionization. It exploits optical coherences to manipulate matter-wave interferences 
to ultimately steer the ionization dynamics into a desired target outcome. In its simplest form, coherent control of photoionization 
is achieved by adjusting the relative phase  
\hbox{\cite{PRLYin92,yuan2016interference,GrumPRA2015,grum2015photoelectron,douguet2016photoelectronCirc,gryzlova2018quantum,DemekhinPRL2018}} or 
time-delay~\cite{wollenhaupt2002interferences,GoetzPRL2019}  
between the frequency components of bichromatic fields promoting single and two-photon ionization. 
The mutual coherence between both frequency components determines the coherence 
properties of the released photoelectron wave packet. These are imprinted in the 
angular distribution of the photoelectron, which has led to its use in the 
control of photoelectron angular distributions (PAD)~\cite{PRLYin92,yuan2016interference,GrumPRA2015,grum2015photoelectron,douguet2016photoelectronCirc,gryzlova2018quantum}. 

An experimental application of control of free-electron wave-packet interferences 
consists in shaping the three-dimensional PAD using phase- and polarization-shaped fields. Additional degrees of freedom can be exploited 
by generalizing the bichromatic scenario to cases comprising a manifold of 
interfering multi\-photon ionization pathways. 
Recent experimental works on resonantly-enhanced multi\-photon ionization of potassium\- atoms using amplitude,
phase- and polarization- shaped pulses, made it possible to Fourier synthesize free-electron wave packets, i.e.,
to engineer photoelectron wave packets with tailored momentum distribution, 
by influencing the mutual coherence among the interfering partial 
wave components originating from the various allowed pathways for multi\-photon 
ionization~\cite{wollenhaupt2013tomographic,Kerbstadt2019,WollenhauptFaradayDiss2011}.

Recalling that these synthesized electron matter waves originate from partial
wave packets coherently interfering during their temporal evolution in the continuum~\cite{wollenhaupt2013tomographic}, 
the question arises whether such interferences can be coherently manipulated to further control
matter-induced processes in a target system.   
 \begin{figure}[!tb]
 \centering
 \includegraphics[width=0.99\linewidth]{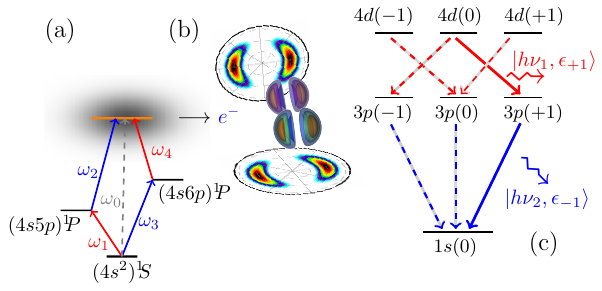}
	\caption{\textbf{Matter-wave control of correlated photon-pair generation:} (a) Ionization of atom~\textit{A} by a classical field probing 
	a manifold of interfering ionization pathways generates a coherent superposition
	of continuum states that result in the photoelectron wave packet with a momentum distribution depicted in~(b). 
	Due to inelastic excitation induced by the electron wave packet upon collision with the target atom~\textit{B}, 
	the optical decay of the latter results in the emission of entangled photon pairs (c).  
}
 \label{fig:figure1}
 \end{figure}

In this Letter, we investigate the capabilities of synthesized free-electron 
wave packets to control dynamical processes in matter-matter interactions. 
As an application, we review and adapt to a modern contextual technological 
framework the pioneering works on coincidence detection measurements of entangled photon-pair
generation by electron bombardment of a target atom~\cite{ClauserPhysRevD1974} and scrutinize the control of entangled photon-pair generation triggered 
by the coherent interaction of  synthesized free-electron wave packets colliding with a target \hbox{atom $B$.} The incident photoelectron originates from coherently
controlled interferometric multi\-photon ionization of its parent atom $A$, as depicted in Fig.~\ref{fig:figure1}(a). It is composed of two partial
wave packets originating from each ionization pathway, as illustrated in Fig.~\ref{fig:figure1}(b). Both components 
propagate under the field-free Hamiltonian while coherently interfering in the continuum as they propagate towards the target. Interaction with the latter
results in elastic scattering of the incoming wave packet and inelastic excitation of the target. We analyze the emission  
of entangled photon pairs in subsequent radiative photo-cascade emission from the excited target atom.                      

Light-driven control of entangled photon-pair states produced by parametric down-conversion  
is a current topic of active research with major impact in 
photonic quantum information sciences~\cite{lu2007demonstration,zhou2013calculating,kwiat1995new,guo2017parametric}.  
Control of spontaneous fluorescence (intensity) emission has been reported in single~\cite{paspalakis1998phase,quang1997coherent,ghafoor2000amplitude,sun2009amplitude} 
and parametric-down photon emission~\cite{jeronimo2009control} using coherent and chirped light sources and by 
exploiting the twist phase of a Gaussian beam with partial transverse spatial coherence~\cite{hutter2020boosting}.  
Here, we exploit the use of synthesized electron matter waves to control the correlated angular distribution  
of entangled photon pairs emission in electron-atom collisions.
\medskip

\paragraph*{Theoretical model.---} 
The atom-photon field inter\-action is treated at the level of the Weisskopf-Wigner theory for spontaneous emission~\cite{weisskopf1997berechnung}. 
De-excitation of the target atom~\textit{B} occurring during and after collision 
with the incident electron wave packet yields the photon field 
in an excited multi\-mode state. The basis,  
\begin{eqnarray}
	\label{eq:photon_basis}
	|n_{\displaystyle{_{\textbf{\text{k}}_1,\boldsymbol{\sigma}_1}}}, n_{\displaystyle{_{\textbf{\text{k}}_2,\boldsymbol{\sigma}_2}}},\dots\rangle 
	\!=\!
\prod_{\displaystyle{_{\textbf{\text{k}}_j,\boldsymbol{\sigma}_j}}}\!\!|n_{\displaystyle{_{\textbf{\text{k}}_j,\boldsymbol{\sigma}_j}}}\rangle ,
\end{eqnarray}
represents $n_{\textbf{\text{k}}_j,\boldsymbol{\sigma}_j}$ photons in mode $(\textbf{\text{k}}_j,\boldsymbol{\sigma}_j)$ with momentum $\hbar\textbf{\text{k}}_j$ 
and polarization $\hat{\boldsymbol{\epsilon}}_{\sigma_j}$ 
subject to the transversality conditions $\hat{\boldsymbol{\epsilon}}_{\boldsymbol{\sigma}_j}\cdot\textbf{\text{k}}_j=0$. Atom $B$ and the photon field are coupled
via the terms $\Op{A}_s(\Op{r})\cdot\Op{p}$ and $\Op{A}_s(\Op{r})\cdot\Op{A}_s(\Op{r})$, with
\begin{eqnarray}
        \label{eq:As} 
	\Op{A}_s(\Op{r})&=&\!\!\!\sum_{\displaystyle{_{\textbf{\text{k}}_j,\boldsymbol{\sigma}_j}}}\!\!\! 
	A_0(k_j)\!\left[\Op{a}_{\displaystyle{_{\textbf{\text{k}}_j,\boldsymbol{\sigma}_j}}}\! e^{i\textbf{\text{k}}_j\cdot\Op{r}}\,\hat{\text{\textbf{e}}}_{\boldsymbol{\sigma}_j} 
	\!+\! 
	\Op{a}^\dagger_{\displaystyle{_{\textbf{\text{k}}_j,\boldsymbol{\sigma}_j}}}\! e^{-i\textbf{\text{k}}_j\cdot\Op{r}}\,\hat{\text{\textbf{e}}}_{\boldsymbol{\sigma}_j} 
	\right] \quad
\end{eqnarray}
as the vector potential operator coupling the eigenstates of~\textit{B} with the photon field. 
$\Op{a}^\dagger_{\textbf{\text{k}}_j,\boldsymbol{\sigma}_j}$ ($\Op{a}_{\textbf{\text{k}}_j,\boldsymbol{\sigma}_j}$) 
creates (annihilates) one photon in mode $(\textbf{\text{k}}_j,\boldsymbol{\sigma}_j)$. 
The Hamiltonian, 
\begin{subequations}
\label{eq:Hamiltonian}
\begin{eqnarray}
\label{eq:H_AB}
	\Op{H}_{AB}(t)&=&\left[ \Op{H}_{A} - e\Op{r}\cdot\Op{E}(\Op{r},t)\right]\otimes \mathbb{1}  +  \Op{V}_{I}\\
	       &+& \!\mathbb{1}\!\otimes \!\left[\sum_{\textbf{\text{k}}_j,\boldsymbol{\sigma}_j}\!\!\hbar\omega(\textbf{\text{k}}_j) + \dfrac{1}{2m}\left(\Op{p}-\frac{e}{c}\Op{A}_s(\Op{r},t)\right)^2\right],\nonumber
\end{eqnarray}
	dictates the ionization dynamics of atom~\textit{A}, scattering
of the resulting photoelectron wave packet by atom~\textit{B}, excitation of the latter due to 
collision, and photoemission upon de-excitation of atom~\textit{B}. 
The interaction,  
\begin{eqnarray}
     \label{eq:V_I}
     \Op{V}_{I} &\!=\!&
V^B_{ne}(\Op{r}\!-\!\boldsymbol{R}_{B})\!\otimes\mathbb{1} +\Op{V}_{\tiny\! ee}(\Op{r},\Op{r}^\prime)  + \mathbb{1}\otimes\! V^B_{ne}(\Op{r}\!-\!\boldsymbol{R}_{B}),\quad\quad
\end{eqnarray}
\end{subequations}
mediates the elastic scattering as well as inelastic excitation of atom~\textit{B} with no change in the distribution of the photon modes. 
Ionization of atom~\textit{A} is 
controlled by the classical field $\boldsymbol{E}(\boldsymbol{r},t)\!=\!\boldsymbol{E}(t)\,f_{\Omega_A}(\boldsymbol{r})$,
with $f_{\Omega_A}(\boldsymbol{r})$ a Heaviside function.  The latter ensures a constant spatial distribution   
in the vicinity of atom~\textit{A} and leaves the target atom~\textit{B} unaffected.
$V^B_{ne}(\Op{r}\!-\!\boldsymbol{R}_{B})$ is the potential energy 
due to the effective nuclear charge distribution of atom~\textit{B} with $\!\boldsymbol{R}_{B}$ the (fixed) origin of the coordinates of~\textit{B}
and $\Op{V}_{\tiny\!ee}(\Op{r},\Op{r}^\prime)$ the potential energy interaction between the incoming electron and the electron in the target. To simplify the notation, Eq.~\eqref{eq:H_AB} is written in
the effective tensor product basis $|\psi^{A}_{\gamma^A_a}\rangle\otimes|\widetilde{\psi}^{B}_{\gamma^B_b}\rangle$,
where $|\widetilde{\Phi}^B_{\gamma_B}\rangle \equiv |\Phi^B_{\gamma_B}\rangle \otimes |n_{\displaystyle{_{\textbf{\text{k}}_1,\boldsymbol{\sigma}_1}}}, 
n_{\displaystyle{_{\textbf{\text{k}}_2,\boldsymbol{\sigma}_2}}},\dots\rangle$, with 
 $|\Phi^A_{\gamma^A_a}\rangle$ and $|\Phi^B_{\gamma^B_b}\rangle$
as the eigenvectors of the isolated Hamiltonians, $\Op{H}_{A}$ and $\Op{H}_{B}$,  of atoms $A$ and $B$, respectively,
satisfying $\small \Op{H}_{A}|\Phi^{A}_{\gamma^{A}_a}\rangle=\epsilon^{A}_{\gamma_{a}}|\Phi^A_{\gamma^A_a}\rangle$ and 
$\small \Op{H}_{B}|\Phi^{B}_{\gamma^B_b}\rangle=\epsilon^{B}_{\gamma_{b}}|\Phi^B_{\gamma^B_b}\rangle$.
With this contracted notation, the first term on the rhs of Eq.~\eqref{eq:H_AB}
applied to an element, e.g., $|\Phi^A_{\gamma_A}\rangle\otimes|\widetilde{\Phi}^B_{\gamma_B}\rangle$, 
will effectively only act on $|\Phi^A_{\gamma_A}\rangle$, leaving the component $|\widetilde{\Phi}^B_{\gamma_B}\rangle$ unaltered.  
Conversely, the third term (second line) in Eq.~\eqref{eq:H_AB} acts on $|\widetilde{\Phi}^B_{\gamma_B}\rangle$, leaving the component $|\Phi^A_{\gamma_A}\rangle$ unchanged. 
The same applies to Eq.~\eqref{eq:V_I}, where $\Op{r}^\prime$ and $\Op{r}$ in $\Op{V}_{\tiny\!ee}(\Op{r},\Op{r}^\prime)$ are used to symbolize a two-electron potential energy operator,
in contrast to the one-electron operators in the first and third terms; see Appendix.

We solve the time-dependent Schr\"odinger equation  
\begin{figure*}[!tb]
\centering
 \includegraphics[width=0.95\linewidth]{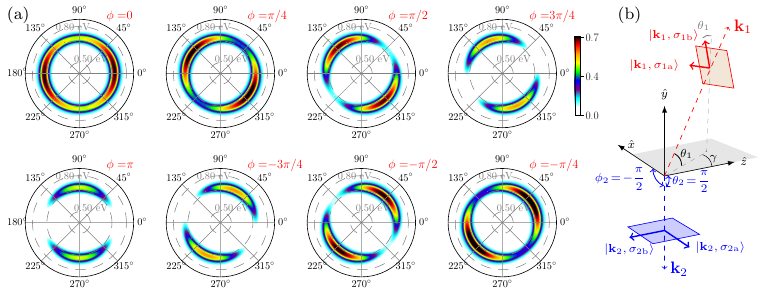}
	\caption{\textbf{Inelastic excitation-decay control mechanism:} (a) Angular distribution 
	of correlated photon-pair emission as a function
        of the relative phase, $\phi$, between the laser frequencies $\omega_1$ and $\omega_3$ 
        shown in Fig.~\ref{fig:figure1}(a). The incident photoelectron wave packet results 
	from a coherent superposition of even-parity two-photon ionization pathways. 
	Note that the angular probability of photodetection follows the relative phase.
	(b) Correlated photon-pair detection scheme 
	discussed in the text ($\gamma=0$).} 
\label{fig:figure2}
\end{figure*}

\begin{subequations}
\begin{eqnarray}
i\dfrac{\partial}{\partial t}|\Psi_S(t)\rangle=\Op{H}_{AB}(t)|\Psi_S(t)\rangle\,,
\end{eqnarray}
and write $|\Psi_S(t)\rangle$ as a coherent superposition in the antisymmetrized tensor product space 
spanned by the eigen\-vectors of the isolated Hamiltonians and Eq.~\eqref{eq:photon_basis}, i.e.,
	\begin{eqnarray}
	\label{eq:Psi_S}
	|\Psi_S(t)\rangle&=&\!\!\! \sum_{\gamma^A_a,\gamma^B_b}\,
	\sum_{n_{\displaystyle{_{\textbf{\text{k}}_1,\boldsymbol{\sigma}_1}}}} 
	\sum_{n_{\displaystyle{_{\textbf{\text{k}}_2,\boldsymbol{\sigma}_2}}}}\!\! \dots\, 
	|\Phi_{\gamma^A_a,\gamma^B_b}\rangle
	\!\otimes\!\,\,\!\!\!\!\prod_{\displaystyle{_{\textbf{\text{k}}_j,\boldsymbol{\sigma}_j}}}\!\!|n_{\displaystyle{_{\textbf{\text{k}}_j,\boldsymbol{\sigma}_j}}}\rangle \nonumber \\[0.1cm]
	&\times& \exp\left[-i\left(\!\epsilon^A_{\gamma_a}\!\!+\!\epsilon^B_{\gamma_b}\!+\!\sum_{\text{\textbf{k}}_i,\boldsymbol{\sigma}_i}\!\omega(\textbf{\text{k}}_i)\,n_{\displaystyle{_{\textbf{\text{k}}_i,\boldsymbol{\sigma}_i}}}\right)t\right]\nonumber\\[0.2cm] 
	&\times& S_{\gamma^A_a,\gamma^B_b}(n_{\displaystyle{_{\textbf{\text{k}}_1,\boldsymbol{\sigma}_1}}}, n_{\displaystyle{_{\textbf{\text{k}}_2,\boldsymbol{\sigma}_2}}},\,n_{\displaystyle{_{\textbf{\text{k}}_3,\boldsymbol{\sigma}_3}}} \dots ;t ).
\end{eqnarray}
\end{subequations}
The vector $|\Phi_{\gamma^A_a,\gamma^B_b}\rangle\!\equiv\!|\Phi^{A}_{\gamma^A_a}\rangle\!\otimes\!|\Phi^{B}_{\gamma^B_b}\rangle\!-\!|\Phi^{B}_{\gamma^B_b}\rangle\!\otimes\!|\Phi^{A}_{\gamma^A_a}\rangle$ 
represents one electron in a spin-orbital state 
	$\langle\boldsymbol{r}_{1};m_{s_1}|\Phi^{A}_{\gamma^A_{a}}\rangle$ of~\textit{A} and the other one 
in $\langle\boldsymbol{r}_2;m_{s_{2}}|\Phi^{B}_{\gamma^B_{b}}\rangle$ of~\textit{B} 
with $m_{s_{1,2}}$ the spin-magnetic quantum numbers for electrons $1$ and $2$. 
Note that 
$\langle\boldsymbol{r}_1,m_{s_{1}};\boldsymbol{r}_2,m_{s_{2}}|\Phi_{\gamma^A_a,\gamma^B_b}\rangle\!=\!-\langle\boldsymbol{r}_2,m_{s_{2}};\boldsymbol{r}_1,m_{s_{1}}|\Phi_{\gamma^A_a,\gamma^B_b}\rangle$.
Summation over $\gamma^A_a$ includes all bound  and
continuum states, and $\omega(\textbf{\text{k}}_j)=|\textbf{\text{k}}_j|\, c$. Matrix elements
to describe the ionization process were obtained with the $B$-spline $R$-matrix codes~\cite{zatsarinny2006bsr}.

Due to the many-body character of the stimulated photon field, a full time-dependent 
treatment of the dynamics, while keeping 
track of every possible photon mode 
emitted and absorbed during the aforementioned processes, 
is a formidable and computationally prohibitive task. 
To scrutinize the 
control mechanisms for radiative photon cascade emission triggered by the collision
while keeping the calculations tractable, we resort to obtain the time-dependent coefficients in Eq.~\eqref{eq:Psi_S}  
from a time-dependent perturbative series expansion up to order~$k$.
Details are provided in the Appendix. 
\medskip

\paragraph*{Inelastic excitation-decay control mechanism.---}
We start by considering resonantly-enhanced two-pathway 
coherent ionization of Ca as schematically
depicted in Fig.~\ref{fig:figure1}(a). Resonant excitation of 
the $(4s5p){}^1P$ and $(4s6p)^1P$ states 
is mediated by the frequency components $\omega_1\!=\!4.554\,$eV and $\omega_3\!=\!5.167\,$eV
of a classical field, both left-circularly polarized. Their relative phase is used as a control parameter. Ionization is ensured by 
the linearly polarized frequency components $\omega_2\!=\!17.324\,$eV and $\omega_4\!=\!16.721\,$eV with a flat spectral phase.
The frequency components have a duration of $20\,$fs (FWHM) and are not time-delayed. Each ionization pathway generates a coherent superposition of continuum eigenstates of~\textit{A}
peaked at the same photoelectron energy of $15.755\,$eV, as depicted in Fig.~\ref{fig:figure1}(b). The 
angular-momentum components defining the wave packets that arise from each of these ionization channels are coherently
combined, carrying the temporal coherence of the classical field. 

The target atom~\textit{B}, initially in its ground state, is taken as the hydrogen atom. 
The quantization axis is defined by the vector parallel to the $z$ direction connecting
both atoms, situated at the positions $\boldsymbol{R}_{A}$ and $\boldsymbol{R}_B$ with $|\boldsymbol{R}_{B}-\boldsymbol{R}_A|=R_{B,z}-R_{A,z}= 1200\,$ atomic units.
After excitation by the tailored electron wave packet, optical decay  
may occur via different de-excitation pathways allowed by selection rules, as epitomized in Fig.~\ref{fig:figure1}(c).

The angular distribution of the emitted photons is obtained using the multipole expansion
in Eq.~\eqref{eq:As},  
\begin{eqnarray}
e^{\pm i\textbf{\text{k}}_{j}\cdot\textbf{\text{r}}}\!=\!4\pi\!\sum_{\lambda,\mu} (\pm i)^{\lambda} j_\lambda(k_jr)Y^{\lambda}_{\mu}(\theta_r,\phi_r)\, Y^{\lambda*}_{\mu}(\theta_{k_j},\phi_{k_j}),\,\quad
\end{eqnarray}
with the spherical harmonics $Y^\lambda_{\mu}(\theta_{k_j},\phi_{k_j})$ for 
the angles of photoemission defining the mode $(\textbf{\text{k}}_j,\boldsymbol{\sigma}_j)$. 
The polarization components 
of the emitted photons are obtained according to  
$\hat{\textbf{\text{e}}}_{\sigma_{ja}}\!\!=\!\textbf{\text{k}}_j\!\times\!\hat{\textbf{\text{e}}}_0/|\textbf{\text{k}}_j\!\times\!\hat{\textbf{\text{e}}}_0|$ 
and $\hat{\textbf{\text{e}}}_{\sigma_{jb}}\!\!=\!\textbf{\text{k}}_j\!\times\!\hat{\textbf{\text{e}}}_{\sigma_{ja}}/ |\textbf{\text{k}}_j\!\times\!\hat{\textbf{\text{e}}}_{\sigma_{ja}}|,$ where 
\begin{eqnarray}
\label{eq:multipole}
\textbf{\text{k}}_j &=& (4\pi/3)^{1/2}\,\, |\textbf{\text{k}}_j|\!\!\sum_{q=0,\pm 1}\!\!\text{Y}^{1}_{q}(\theta_{_{k_j}},\phi_{_{k_j}} )\,\hat{\textbf{\text{e}}}^{*}_q, 
\end{eqnarray}
with $\hat{\textbf{\text{e}}}_q$ the covariant spherical unit vectors. Both polarization vectors
are functions of the angles $(\theta_{k_j},\phi_{k_j})$.

The coincidence photodetection scheme is shown in 
Fig.~\ref{fig:figure2}(b): a first photodetector, 
fixed at $\theta_{k_2}\!\!=\!\!\pi/2$ and $\phi_{k_2}\!=\!-\pi/2$, measures 
the polarization component along the \hbox{$z$-axis} in Fig.~\ref{fig:figure2}(b) 
of a photon with energy $h\nu_2\!=\!12.078\,$eV, 
corresponding to the radiative transition $3\text{p}(m\!=\!\pm 1,0)\!\rightarrow \! 1\text{s}$ in Fig.~\ref{fig:figure1}(c). The state of such photon is hereafter referred to as mode (2). 

A second detector, fixed at  $\phi_{k_1}\!=\!\pi/2$ but free to move along the polar coordinate $\theta_{k_1}$, scans, along $\theta_{k_1}$, the direction of emission of the entangled peer
defined by the state $|\textbf{\text{k}}_1,\sigma_{1b}\rangle$ in Fig.~\ref{fig:figure2}(b): 
a photon of energy $h\nu_1=0.661\,$ eV, corresponding to the radiative transition $4d(\{m\})\rightarrow 3p(\pm 1,0)$   
polarized along $\hat{\textbf{\text{e}}}_{\sigma_{1b}}$. 

\begin{figure}[!tb]
\centering
\includegraphics[width=0.93\linewidth]{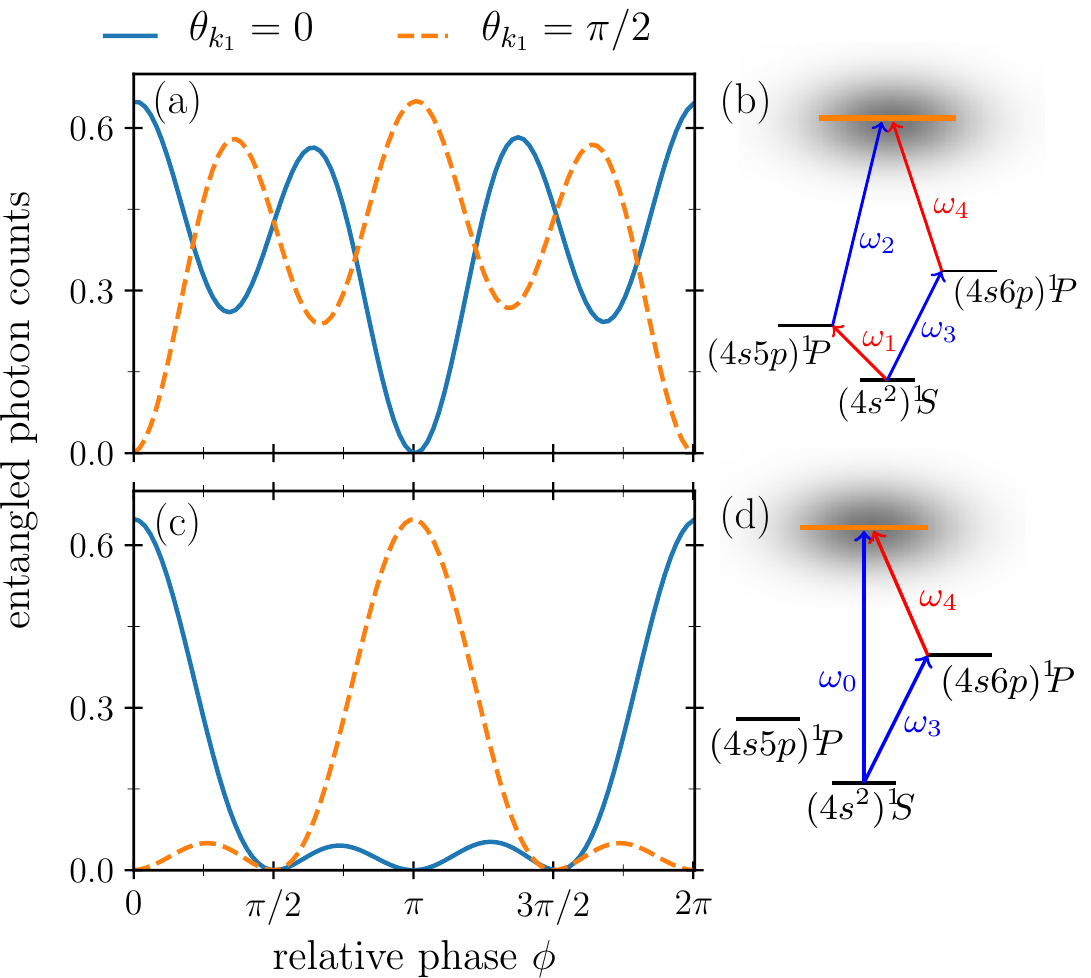}
	\caption{ (a) Photodetection probability at angles $\theta_{k_{1}}=0$ (solid blue line)
	and $\theta_{k_1} =\pi/2$ (dashed orange line)
	as a function of the relative phase between
	the laser frequencies $\omega_1$ and $\omega_3$ shown in (b)
	when a photon in mode (2) is simultaneously detected.
	The incident electron wave packet originates from 
	the ionization scheme shown in (b). (c) Same as (a) but as a function
	of the relative phase between the laser frequencies $\omega_0$ and $\omega_3$
	shown in~(d).} 
\label{fig:figure3}
\end{figure}

Figure~\ref{fig:figure2}(a) shows the angular probability distribution of measuring
the correlated photon in coincidence with its entangled peer in mode (2)
as a function of the relative phase between the frequencies $\omega_1$ and $\omega_3$ 
of Fig.~\ref{fig:figure1}(a). The direction of emission exhibits 
a noticeable dependence on the temporal coherence conveyed by the incident photoelectron wave packet: 
the probability of entangled photon-pair detection
is strongly affected by the mutual phase of the photoionization probability amplitudes, controlled by the relative
phase between the frequency components of the classical field 
probing the contributing photoionization pathways. 

The angle-resolved occurrence of coincident photo\-detection 
is also sensitive to the parity of the photoionization pathways probed to engineer the incident 
photoelectron wave packet. This is shown in Fig.~\ref{fig:figure3},  
comparing, at the fixed emission angles $\theta_{k_1}=0$ and $\theta_{k_1}=\pi/2$, 
the probability of coincident photodetection
already discussed in Fig.~\ref{fig:figure2}, this time using different photoionization schemes
to engineer the incident electron wave packet:  
same- and opposite-parity photoionization pathways.

Figure~\ref{fig:figure3}(a) displays the probability of coincident photodetection 
obtained when the incident photo\-electron wave packet is engineered
according to the resonantly-enhanced two-photon ionization scheme
promoting even-parity pathways depicted in Fig.~\ref{fig:figure3}(b). As shown in Fig.~\ref{fig:figure3}(a), 
the probability for simultaneous photon-pair 
detection at a given direction $\theta_{k_1}$ can be entirely suppressed or 
enhanced depending on the relative phase between the contributing photoionization pathways. 
\begin{figure}[!tb]
\centering
\includegraphics[width=0.80\linewidth]{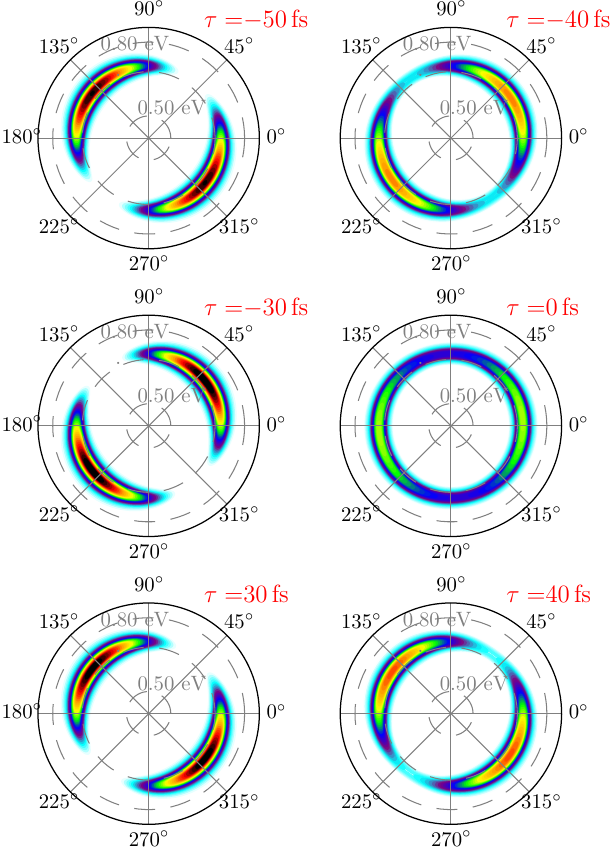}
	\caption{ Time-resolved probability for coincident photo\-detection  
	as a function of the time delay between the pump and probe pulses discussed in the text. Colormap as in Fig.~\ref{fig:figure2}.}
\label{fig:figure4}
\end{figure}

Likewise, as shown in Fig.~\ref{fig:figure3}(c), opposite-parity photoionization pathways, as depicted in Fig.~\ref{fig:figure3}(d),
can also be exploited to engineer the photoelectron wave packet to ultimately suppress 
or enhance the probability of correlated photon-pair detection. 
In this case, control is achieved by manipulating the relative phase between 
the one- and two-photon ionization
pathways  
through the rela\-tive phase between the frequencies $\omega_0$ 
and $\omega_3$. 

As defined,  the relative phase corresponding to $\phi=0$ maximizes (minimizes)
the probability of detection in the direction $\theta_{k_1}=0$ ($\theta_{k_1}=\pi/2$)
for both photoionization schemes depicted in Figs.~\ref{fig:figure3}(b) and (d)
when a photon in mode (2) is simultaneously detected. Conversely, for $\phi=\pi$,  
the probability of detection in the direction $\theta_{k_1}=0$ ($\theta_{k_1}=\pi/2$)
is minimized (maximized) for both schemes. In contrast, the relative phases corresponding to $\phi=\pi/2$ and $\phi=3\pi/2$ result in
the suppression of entangled photon-pair coincident detection at angles $\theta_{k_1}=0$ and  
$\theta_{k_1}=\pi/2$ for the case of an odd-even parity photoionization pathway (cf.\ Fig.~\ref{fig:figure3}(c)),
whereas its even-parity counterpart 
results in an equal probability of coincidence photodetection, cf.~Fig.~\ref{fig:figure3}(a).

Control of directional correlated photon-pair detection can also be achieved by adjusting the relative time delays between the different 
pulses carrying the various frequencies components in Fig.~\ref{fig:figure1}(a). To illustrate this,   
we choose two pump-laser frequencies, $\omega_1$ and $\omega_3$, to resonantly excite the $(4s5p)^1P$ and $(4s6p)^1P$ states in atom~\textit{A}, 
creating a superposition of states evolving according to the free-field Hamiltonian. After a delay $\tau$, 
a probe field with frequencies $\omega_2$ and $\omega_4$ is introduced, ionizing the electron in the coherent superposition. 
The resulting photoelectron wave packet then carries the coherence of the superposition of bound states, defined by the phase accumulated  
between the pump and probe pulses. Figu\-re~\ref{fig:figure4} shows the time-resolved
probability  of correlated photon-pair detection as a function of the delay $\tau$.  
The probability for coincident photodetection upon sequential de-excitation of atom $B$, located far from atom $A$, 
is sensitive to the phase of the coherent superposition state in atom $A$, carried by the colliding photoelectron wave packet. For a fixed direction $\theta_{k_1}$, the photon yield
can be controlled significantly.  Compare, for example, the yields at $\theta_{k_1}=45^\circ$ 
 for the delays $\tau=-50\,$fs and $\tau=-40\,$fs. 

The positions of the maxima and minima in Fig.~\ref{fig:figure2}(a) are dictated
by the transversality condition for the mode $|\textbf{\text{k}}_1,\sigma_{1b}\rangle$ and
the magnetic quantum number $m$ of the $4d(m)$ state involved in the de-excitation transition to the $3p(0)$ state. 
The photon with energy $h\nu_1$ is linearly polarized, i.e., parallel to the $\hat{z}$~axis in Fig.~\ref{fig:figure2}(a)),
if the transition $4d(0)\rightarrow 3p(0)$ takes place. In this case, the maxima correspond to $\theta_{k_1}=90^\circ$ and $\theta_{k_1}=270^\circ$, since 
$\sigma_{1b}$  is along the $\hat{z}$~direction. The minima correspond to the angles $\theta_{k_1}=0^\circ$ and $\theta_{k_1}=180^\circ$,
since $\sigma_{1b}$ is then perpendicular to the $\hat{z}-$ axis of polarization: the detector 
counting photons polarized along $\sigma_{1b}$, then parallel to the $y-$ axis, finds no signal. 
Conversely, if the transition $4d(m^\prime)\rightarrow 3p(0)$, with $m^\prime=\pm 1$, takes place, then the photon it is circularly polarized,
and the aforementioned positions for the maxima and minima are inverted. By detecting a photon in mode (2) alone, 
it is not possible to determine the pathway taken by the photon of energy $h\nu_1$ in the first step of the cascade.
For an arbitrary $\theta_{k_1}$, several de-excitation pathways may contribute to the detected polarization $\sigma_{1b}$ if the $4d(m)$ states are populated.
The position of maxima and minima then fluctuates between the limiting values obtained when only one $4d(m)$ state is populated at a time. 
The angular distribution of the photons with energy $h\nu_1$ contains coherent contributions from each de-excitation pathway. 
These depend on the population and phase of the $4d(m)$ states excited via coherent inelastic excitation preceding the radiative decay.
Control of the angular distribution is achieved if different de-excitation pathways  
contribute to the same emitted photon mode. 
Their mutual coherence can be  adjusted by controlling the ionization process generating the incident photoelectron wave packet. 

\paragraph*{Conclusions.---}

Motivated by the recent developments in free-electron wave packet interferometry 
and engineering, and the increasingly active research in 
coherent control of entangled photon-pair generation, 
we investigated the generation and control of correlated photon pairs triggered by electron-atom collisions.
In contrast to standard approaches, we demonstrated the possibility of 
controlling quantum light by employing engineered matter waves. Using calcium and hydrogen 
as prototypes to control entangled two-photon cascade emission 
triggered by electron-atom collisions, we demonstrated that 
laser-synthesized free-electron matter waves can be used to  coherently control processes in  
matter-matter interactions. Taking advantage of coherently-driven electron-electron interactions to induce, e.g., population transfers may be beneficial when the efficiency to accomplish 
the same task using light sources is restricted by electric dipole selection rules. 
Our results can be extended to more complex cases, such as chiral molecules, with
the potential to reveal new insights into the interaction of quantum light with chiral
matter waves. We also foresee extending our approach to electron-ion collisions to
investigate the control of correlated photon-pair emission mediated by electron-trapping
correlated decay. 

This work was supported by the National Science Foundation under grant No.\ \hbox{PHY-1803844}.

\section*{Appendix}
\appendix

\section{\textsc{Hamiltonian system}}
\label{sec:section1}
\subsection{\textsc{Isolated Hamiltonians}}
In order to keep the calculations tractable,                                                                             
we consider two initially isolated, non-interacting atomic systems, labelled $A$ and $B$,
with Hamiltonians                                                                                               
\begin{subequations}
   \label{eq:hamiltonians}
   \begin{eqnarray}
      \label{eq:hamiltonian_A}
	   \Op{H}_A&\!=\!&\sum^{N_A}_{i=1}\left(\dfrac{\Op{p}^2_i}{2m}+V^A_{ne}(\Op{r}_i-\Op{R}_{A})+\!\sum^{N_A}_{i^\prime>i} V_{ee}(\Op{r}_i-\Op{r}_{i^\prime})\!\right)\!,\quad\quad\\[-0.5cm]\nonumber
   \end{eqnarray}
corresponding to that of the atom from which the photoelectron wave packet is released, and
   \begin{eqnarray}
      \label{eq:hamiltonian_B}
       \Op{H}_B&\!=\!&\sum^{N_B}_{j=1}\left(\dfrac{\Op{p}^2_j}{2m}+V^B_{ne}(\Op{r}_j-\Op{R}_{B})+\!\!\sum^{N_B}_{j^\prime>j}\!V_{ee}(\Op{r}_j-\Op{r}_{j^\prime})\!\right)\quad\quad 
\end{eqnarray}
\end{subequations}
for the target atom. They fulfill $\Op{H}_A|\Phi^{A}_{\gamma^A_a}\rangle=\epsilon^{A}_{\gamma_{a}}|\Phi^A_{\gamma^A_a}\rangle$ 
and $\Op{H}_B|\Phi^{B}_{\gamma^B_b}\rangle=\epsilon^{B}_{\gamma^B_{b}}|\Phi^B_{\gamma^B_b}\rangle$ with
eigenvalues $\epsilon^A_{\gamma_a}$ ($\epsilon^B_{\gamma_b}$),
where $\gamma^A_a$ ($\gamma^B_b$) collectively denotes a set of quantum numbers that uniquely define 
the states $|\Phi^{A}_{\gamma^A_a}\rangle$ and $|\Phi^{B}_{\gamma^B_b}\rangle$. 
The summations are over the $N_A$ and $N_B$ electrons of atom $A$ and $B$, respectively. 
$\Op{V}_{ee}$ is the electron-electron potential energy operator  
and $\Op{R}_A$ ($\Op{R}_B$) the position operator acting on the nuclear wavefunction of $A$~($B$). 

Recoil is not considered. The origins of the coordinate systems are $\boldsymbol{R}_{0,A}$ ($\boldsymbol{R}_{0,B}$),  
defined by the (fixed) position of the point-like nuclear charge distribution assumed 
to be of the form $Z_{A}\,\delta(\boldsymbol{R}\!-\!\boldsymbol{R}_{A})$  
($Z_{B}\,\delta(\boldsymbol{R}\!-\!\boldsymbol{R}_{B})$) with $Z_A$ ($Z_B$) as the
effective nuclear charge. The distance $|\boldsymbol{R}_{A}\!-\!\boldsymbol{R}_{B}|$ 
is taken such that 
$\langle\Phi^A_{\gamma^A_a}|\Phi^B_{\gamma^B_b}\rangle\!=\!0$ for all bound states considered. 
For the numerical calculations, we have set  $|\boldsymbol{R}_{B}-\boldsymbol{R}_A|=R_{B,z}-R_{A,z}= 1200\,$ atomic units.
This axis of quantization defines the frame for the electric field polarization given in Sec.~\ref{sec:optical_preparation}.
We denote by $\mathcal{H}_{A}$ and $\mathcal{H}_{B}$ the Hilbert 
spaces spanned by the eigen\-vectors of the Hamiltonians defined in Eqs.~\eqref{eq:hamiltonian_A} and~\eqref{eq:hamiltonian_B}, respectively. 

\subsection{\textsc{Optical preparation and laser-induced ionization}}
\label{sec:optical_preparation}
The optical preparation and ionization of atom $A$ is mediated by a classical field $\textbf{\text{E}}(\boldsymbol{r},t)$ 
parameterized as 
\begin{eqnarray}
    \label{eq:field} 
	\text{\textbf{E}}(\boldsymbol{r},t)&\!=\!&
	\!\sum_{\mu_0} 
	\sum_{n=1}^N h^{(n)}_{\mu_0}(t-\tau_n) 
	  \,\mathrm{Re}\left\{\!e^{-i\Phi_n(t-\tau_n)}\,\mathbf{e}_{\mu_0}\!\right\}\! f_{\Omega_A}(r),\, \nonumber\\[-0.3cm]
  \end{eqnarray}
with $N$ frequency components $\omega_n$ 
and instantaneous frequencies $\Phi_n(t)\!=\!\omega_n t +\tilde\varphi_n$, 
where $\tilde{\varphi}_n\!=\!-\omega_n\tau_n+\phi_n$
is the spectral phase, $\phi_n$ the carrier envelope phase (CEP), and $\tau_n$ the time delay with respect 
to $t\!=\!0$. $h^{(n)}(t-\tau_n)$ is a Gaussian function with adjustable amplitude centered around~$\tau_n$. Left-circular   
\hbox{($\mathbf{e}_{-1}\!=\!(\mathbf{e}_{x}\!-\!i\mathbf{e}_y)/\sqrt{2}$)}, 
right-circular \hbox{($\mathbf{e}_{+1}\!=\!-(\mathbf{e}_{x}+i\mathbf{e}_y)/\sqrt{2}$)}, and linear \hbox{($\mathbf{e}_{0}\!=\!\mathbf{e}_z$)}
polarization states are described by ${\textbf{e}}_{\mu_0}$. 
The photo\-electron wave packet resulting from ionization 
of~$A$ is controlled by means of these field parameters. 

To avoid treating the collision dynamics in the presence of a dressing background field, the Heaviside function $f_{\Omega_A}(r)$
ensures a constant amplitude for the spatial distribution of~$\textbf{E}$ 
within the region defined by the extension of the outermost excited state of atom~$A$.  
Being zero elsewhere, the target atom~$B$ is not affected by the field. 

\section{Equations of motion}
\label{sec:section2}
\subsection{Perturbation expansion}
To illustrate our idea of coherent control at a reduced computational cost, we 
treat the ionization of atom~$A $ (here calcium) and the collision  
of the resulting photoelectron wave packet with the target atom~$B$ (here atomic hydrogen)
as an effective two-electron problem.
The Hamiltonian, 
\begin{eqnarray}
\label{eq:Hamiltonian}
\label{eq:H_AB}
	\Op{H}_{AB}(t)&=&\left[ \Op{H}_{A} - e\,\Op{r}\cdot\Op{E}(\Op{r},t)\right]\otimes \mathbb{1}  +  \Op{V}_{I}\\
	       &+& \!\mathbb{1}\!\otimes \!\left[\sum_{\textbf{\text{k}}_j,\boldsymbol{\sigma}_j}\!\!\hbar\omega(\textbf{\text{k}}_i) + \dfrac{1}{2m}\left(\Op{p}-\frac{e}{c}\Op{A}_s(\Op{r},t)\right)^2\right],\nonumber
\end{eqnarray}
with $\Op{A}_s$ the vector potential describing the photon field, dictates the ionization dynamics of atom~$A$, scattering
of the resulting photoelectron wave packet by atom~\textit{B}, collision-induced excitation of the latter,
and photoemission upon de-excitation of atom~\textit{B}. 
The interaction $\Op{V}_I$, mediating elastic scattering from as well as 
excitation of atom~\textit{B} without change in the distribution of photon modes  
is defined in Eq.~(3b) in the manuscript. 

We search for a solution, $|\Psi_{S}(t)\rangle$, satisfying  
\begin{subequations}
\begin{eqnarray}
	\label{eq:TDSE}
i\dfrac{\partial}{\partial t}|\Psi_S(t)\rangle=\Op{H}_{AB}(t)|\Psi_S(t)\rangle\,.
\end{eqnarray}
To keep the calculations tractable, we employ a 
	time-dependent perturbative series expansion for 
\begin{eqnarray}
	|\Psi_S(t)\rangle&\approx& |\Psi^{(0)}_S(t)\rangle+\sum^{k_{\text{max}}}_{k=1} |\Psi^{(k)}_S(t)\rangle. 
\end{eqnarray}
The zeroth-order term, $|\Psi^{(0)}_S(t)\rangle$, describes the isolated atoms $A$ and $B$ in their respective ground states $|\Phi^A_{\gamma^{0}_a}\rangle$ 
and $|\Phi^B_{\gamma^{0}_b}\rangle$ and the photon field in the vacuum state 
\begin{eqnarray}
	|\Psi^{(0)}_S(t)\rangle&\!=\!& e^{-i\big(\epsilon^A_{\gamma^{0}_a}+\epsilon^B_{\gamma^{0}_b}+1/2\big) t}\\ 
	&&\quad\times\left[\Phi^A_{\gamma^0_a}\rangle\otimes|\Phi^B_{\gamma^0_b}\rangle-|\Phi^B_{\gamma^0_b}\rangle\otimes|\Phi^A_{\gamma^0_a}\rangle\right]\!\otimes\!|0\rangle.\nonumber 
\end{eqnarray}
At intermediate times, $|\Psi_S(t)\rangle$ is written as a coherent superposition in the antisymmetrized tensor product space 
spanned by the eigen\-vectors of the isolated Hamilto\-nians and a coherent multimode photon field, 
\begin{eqnarray}
\label{eq:Psi_S_k}
	|\Psi^{(k+1)}_S(t)\rangle&=&\!\!\! \sum_{\gamma^A_a,\gamma^B_b}\,
\sum_{n_{\displaystyle{_{\textbf{\text{k}}_1,\boldsymbol{\sigma}_1}}}} 
\sum_{n_{\displaystyle{_{\textbf{\text{k}}_2,\boldsymbol{\sigma}_2}}}}\!\!\! \dots\, 
|\Phi_{\gamma^A_a,\gamma^B_b}\rangle
	\!\otimes\!\,\,\!\!\!\!\prod_{\displaystyle{_{\textbf{\text{k}}_j,\boldsymbol{\sigma}_j}}}\!\!|n_{\displaystyle{_{\textbf{\text{k}}_j,\boldsymbol{\sigma}_j}}}\rangle\nonumber\\[0.1cm]
	&\times& \!\exp\!\left[-i\left(\!\epsilon^A_{\gamma_a}\!\!+\!\epsilon^B_{\gamma_b}\!+\!\sum_{\text{\textbf{k}}_i,\boldsymbol{\sigma}_i}\!\omega(\textbf{\text{k}}_i)\,n_{\displaystyle{_{\textbf{\text{k}}_i,\boldsymbol{\sigma}_i}}}\right)t\right]\nonumber\\[0.2cm] 
	&\times& S^{(k+1)}_{\gamma^A_a,\gamma^B_b}(n_{\displaystyle{_{\textbf{\text{k}}_1,\boldsymbol{\sigma}_1}}}, n_{\displaystyle{_{\textbf{\text{k}}_2,\boldsymbol{\sigma}_2}}},\,n_{\displaystyle{_{\textbf{\text{k}}_3,\boldsymbol{\sigma}_3}}} \dots ;t),
\end{eqnarray}
\end{subequations}
where $|\Phi_{\gamma^A_a,\gamma^B_b}\rangle\equiv|\Phi^{A}_{\gamma^A_a}\rangle\otimes|\Phi^{B}_{\gamma^B_b}\rangle-|\Phi^{B}_{\gamma^B_b}\rangle\otimes|\Phi^{A}_{\gamma^A_a}\rangle.$

\subsection{Observables}
The expansion coefficients in Eq.~\eqref{eq:Psi_S_k} are used to evalu\-ate
the angle-resolved correlated probability of photon-pair emission according to
\begin{eqnarray}
	\label{eq:observable}
	\dfrac{d^2\sigma_{n_f}}{d(h\nu_{1})\,d\Omega_{\hat{k}_1}} &=&\int d^3p^A_a\,\, 
	\left|\sum^{k_{\text{max}}}_{k=0} S^{(k)}_{\boldsymbol{p}^A_a,\gamma^{B}_{b^0}}(\boldsymbol{n}_{f};t\rightarrow\infty)\right|^2\!,~~~
\end{eqnarray}
where the incoherent summation (integration) is performed over the momenta of the scattered photoelectron. Ionization of atom $B$ by electron impact is not considered. 
Thus, at $t\!\rightarrow\!\infty$, atom $B$ 
returns to its ground state defined by the quantum numbers $\gamma^B_{b^0}$. At intermediate times, the photon field may be excited 
following complex emission and absorption dynamics.  The photon-field configuration  $\boldsymbol{n}_f$ in Eq.~\eqref{eq:observable} specifies  
the distribution of the photon modes of interest at $t\rightarrow\infty$. 

The photon-field configuration corresponding to the radiative 
two-photon cascade emission depicted in Figs.~(1c) and~(2b) in the manuscript   
corresponds to the distribution of modes defined by $\textbf{n}_f=[0,0,...,1_{m_1},...,0, ...,1_{m_2},0,...]^{\text{\sffamily{T}}}$,
where $m_2$ denotes the mode of photon (2). It is defined by the photon energy $h\nu_2=12.078\,$eV, polarization component along $\sigma_{2a}$ in Fig.~(2b),
and direction of photodetection defined by the (fixed) angles $\theta_{k_2}\!=\!\pi/2$ and $\phi_{k_2}\!=\!-\pi/2$ in Fig.~(2b). For additional details, see Sec.~\ref{subsubsec:application_corr_scheme} below. 
On the other hand, $m_1$ denotes the photon mode defined by $h\nu_1=0.661\,$eV, polarization
component along $\sigma_{1b}(\theta_{k_1})$ in Fig.~(2b), as the photon in mode~(1) is detected
as a function of the angle $\theta_{k_1}$, cf.~Sect.~\ref{subsubsec:application_corr_scheme}.

\subsection{Propagation of the time-dependent coefficients}
\label{subsec:convention}
To alleviate the notation, we introduce the occupation number representation
$\boldsymbol{n}=[n_1, n_2, \, ...\, n_p\, ...]$ describing $n_p$ photons in mode $p$. 
Analogously, we define $\boldsymbol{n}\pm\boldsymbol{1}_p=[n_1, n_2, \, ...\, n_p\pm 1\, ...]$
and $\boldsymbol{n}\pm\boldsymbol{2}_p=[n_1, n_2, \, ...\, n_p\pm 2\, ...]$ 
and $|\boldsymbol{n}\rangle=|n_1, n_2, ...,n_p, ...\rangle$.  The indices run
over all photon modes, hereafter referred to as $\Omega_{\gamma}$.

The time-dependent expansion coefficients in Eq.~\eqref{eq:Psi_S_k} are obtained
by projecting Eq.~\eqref{eq:TDSE} onto $|\Phi^{A}_{\gamma^{\prime A}_a}\Phi^{B}_{\gamma^{\prime B}_b};\boldsymbol{n}^\prime\rangle$
and iteratively solving the recursive equation
\begin{subequations}
\begin{widetext}
\begin{eqnarray}
	\label{eq:S1}
S^{(k+1)}_{\gamma^{\prime A}_a,\gamma^{\prime B}_b}(\boldsymbol{n}^\prime;t)&=&
	\int^{t}_{-\infty} \langle\Phi^{A}_{\gamma^{\prime A}_a}\Phi^{B}_{\gamma^{\prime B}_b};\boldsymbol{n}^\prime|\tilde\Psi^{(k)}(t^\prime)\rangle dt^\prime
	+\!\!\!\sum_{\!\gamma^A_a,\gamma^B_b}\!\left[S^{(k)}_{\gamma^{A}_a,\gamma^{B}_b}(\boldsymbol{n}^\prime;t)
	\,e^{(\epsilon^{A}_{\gamma^\prime_a}+\epsilon^{B}_{\gamma^\prime_b}-\epsilon^{A}_{\gamma_a}-\epsilon^{B}_{\gamma_b})t}	
	\right]^{t}_{t_0}\,
	\langle\Phi^A_{\gamma^{\prime A}_a}|\Phi^B_{\gamma^{B}_b}\rangle\langle\Phi^B_{\gamma^{\prime B}_b}|\Phi^A_{\gamma^A_a}\rangle\,,\nonumber\\ 
\end{eqnarray}
\end{widetext}
	with $t_0\rightarrow\!-\infty$. The integrand in Eq.~\eqref{eq:S1} reads 
\begin{eqnarray}
	\label{eq:S2}
\langle\Phi^{A}_{\gamma^{\prime A}_a}\Phi^{B}_{\gamma^{\prime B}_b};\boldsymbol{n}^\prime|\tilde\Psi^{(k)}(t^\prime)\rangle&=&
	\prod_{s\in\Omega_\gamma}\, e^{i(\epsilon^{A}_{\gamma^\prime a}+\epsilon^{B}_{\gamma^\prime b}+\hbar n^\prime_s\omega_s) t^\prime}\quad\\
	&\times&\!\!\langle\Phi^{A}_{\gamma^{\prime A}_a}\Phi^{B}_{\gamma^{\prime B}_b};\boldsymbol{n}^\prime|\Op{H}_{I}(t^\prime)|\Psi^{(k)}_{S}(t^\prime)\rangle.\!\!\nonumber
\end{eqnarray}
Here we have defined, upon expansion of the quadratic term in  Eq.~\eqref{eq:H_AB},
\begin{eqnarray}
\label{eq:H_I}
	\Op{H}_{I}(t)&=&\left[-e\,\Op{r}\cdot\Op{E}(\Op{r},t) +  
	V^B_{ne}(\Op{r}\!-\!\boldsymbol{R}_{B})\right]\otimes\mathbb{1}+\Op{V}_{\tiny\!ee}(\Op{r},\Op{r}^\prime)\nonumber\\
&&+\,\mathbb{1}\!\otimes\!\left[-\dfrac{e}{mc}\Op{A}_s(\Op{r})\cdot\Op{p}+\frac{e^2}{2mc}\Op{A}_s(\Op{r})\cdot\Op{A}_s(\Op{r})\right]\!\!,\quad\quad	
\end{eqnarray}
	with $V^B_{ne}(\Op{r}\!-\!\boldsymbol{R}_{B})\otimes\mathbb{1}$ the potential-energy interaction between the incident electron 
and the nuclear charge distribution of atom $B$, and  $\Op{V}_{\tiny\!ee}(\Op{r},\Op{r}^\prime)$  the potential-energy 
interaction between the incident and the target electron.	
The second term in  Eq.~\eqref{eq:S1}, proportional to $\langle\Phi^A_{\gamma^{\prime A}_a}|\Phi^B_{\gamma^{B}_b}\rangle\langle\Phi^B_{\gamma^{\prime B}_b}|\Phi^A_{\gamma^A_a}\rangle$, 
arises from the antisymmetrization of the two-electron state vector in Eq.~\eqref{eq:Psi_S_k} and may not vanish due to the overlap between the scattering states of~$A$ and the bound states of~$B$. 
	
After using Eqs.~\eqref{eq:S1},~\eqref{eq:S2}, \eqref{eq:H_AB}, and carrying out some straight\-forward algebra, the expression for the expansion coefficients can be recast into the form 
\begin{widetext}
\begin{eqnarray}
\label{eq:contributions}
S^{(k+1)}_{\gamma^{\prime A}_a,\gamma^{\prime B}_b}(\boldsymbol{n}^\prime;t)&=&
	C^{(k+1)}_{[E]}(\gamma^{\prime A}_a,\gamma^{\prime B}_b, \boldsymbol{n}^\prime;t) 
	+C^{(k+1)}_{[V_B]}(\gamma^{\prime A}_a,\gamma^{\prime B}_b, \boldsymbol{n}^\prime;t) 
	+C^{(k+1)}_{[V_{AB}]}(\gamma^{\prime A}_a,\gamma^{\prime B}_b, \boldsymbol{n}^\prime;t)\nonumber\\[0.15cm] 
&+&
	C^{(k+1)}_{[h\nu]}(\gamma^{\prime A}_a,\gamma^{\prime B}_b, \boldsymbol{n}^\prime;t)
	+C^{(k+1)}_{[X]}(\gamma^{\prime A}_a,\gamma^{\prime B}_b, \boldsymbol{n}^\prime;t).
\end{eqnarray}
\end{widetext}
\end{subequations}
\subsection{Coherent electron-impact dynamics}
Defining $\boldsymbol{\gamma}^\prime\equiv(\gamma^{\prime A}_a,\gamma^{\prime B}_b)$, the first term in Eq.~\eqref{eq:contributions}, 
\begin{eqnarray}
\label{eq:C_E}
C^{(k+1)}_{[E]}\!(\boldsymbol{\gamma}^\prime,\boldsymbol{n}^\prime;t)&\!=\!&ie 
\!\!\sum_{\mu_0,\gamma^A_a}\langle\Phi^A_{\gamma^{\prime A}_a}|\Op{r}_{\mu_0}f_{\Omega_A}(\Op{r})|\Phi^A_{\gamma_a}\rangle\!\!\!\!\!\\[0.05cm]
&\!\times\!&\!\int^{t}_{-\infty}\!\!\!\!\!\! 
S^{(k)}_{\gamma^{A}_a,\gamma^{\prime B}_b}(\boldsymbol{n}^\prime ;t^\prime) e^{i(\epsilon^{A}_{\gamma^\prime_a}-\epsilon^{A}_{\gamma_a}\!) t^\prime}\!\text{E}_{\mu_0}(t^\prime) dt^\prime\!,\!\!\!\!\nonumber
\end{eqnarray}
accounts for the correction to the expansion coefficients due to the classical field defined in Eq.~\eqref{eq:field}. The second term,
\begin{eqnarray}
	\label{eq:C_VB}
	C^{(k+1)}_{[V_B]}(\boldsymbol{\gamma}^\prime,\boldsymbol{n}^\prime ;t ) &=&-i
	\sum_{\gamma^{A}_a}\langle\Phi^{A}_{\gamma^{\prime A}_a}|\Op{V}_{ne}(\Op{r}\!-\!\boldsymbol{R}_{B})|\Phi^A_{\gamma_a}\rangle\!\!\!\!\!\!\!\nonumber\\
&\times&\int^{t}_{-\infty}\!\!\! 
S^{(k)}_{\gamma^{A}_a,\gamma^{\prime B}_b}(\boldsymbol{n}^\prime ;t^\prime ) e^{i(\epsilon^{A}_{\gamma^\prime_a} -\epsilon^{A}_{\gamma_a}) t^\prime}\, dt^\prime, \quad\quad
\end{eqnarray}
accounts for the correction to the scattering component $\gamma^{\prime A}_a$ of the expansion coefficients due to the potential-energy interaction between the 
incoming electron and the nuclear charge density at $r_{0,B}$ while leaving the component $\gamma^{\prime B}_b$ and the photon field unchanged. Next,
\begin{eqnarray}
	\label{eq:C_V12}
	C^{(k+1)}_{[V_{AB}]}(\boldsymbol{\gamma}^\prime,\boldsymbol{n}^\prime ;t ) &=&-i
\!\!\!\sum_{\gamma^A_a,\gamma^B_b}\langle\Phi^A_{\gamma^{\prime A}_a}\Phi^B_{\gamma^{\prime B}_b}|\Op{V}_{\tiny\!ee}(\Op{r},\Op{r}^\prime)|\Phi^A_{\gamma_a}\Phi^B_{\gamma_b}\rangle\nonumber\\[0.1cm]
&\times&\!\!\int^{t}_{-\infty}\!\!\!\!\! 
	S^{(k)}_{\gamma^{A}_a,\gamma^{B}_b}(\boldsymbol{n}^\prime ;t^\prime )\,e^{i\varphi(\boldsymbol{\gamma}^\prime,\gamma^A_a,\gamma^B_b)t^\prime}\! dt^\prime\!,\!\,\,\quad\quad
\end{eqnarray}
where $\varphi(\boldsymbol{\gamma}^\prime,\gamma^A_a,\gamma^B_b)\!\equiv\!\epsilon^{A}_{\gamma^\prime_a}+\epsilon^{B}_{\gamma^\prime_b}-\epsilon^{A}_{\gamma_a}-\epsilon^{B}_{\gamma_b}$, accounts for the interaction between the incoming electron and
the electron initially in atom~$B$. More precisely, Eq.~\eqref{eq:C_V12} accounts 
for the elastic and inelastic excitation of atom $B$ triggered by the incident wave packet. It describes, to lowest order,
the scattering of an initial continuum-state component along $|\Phi^A_{\gamma^A_a}\rangle$ to $|\Phi^A_{\gamma^{\prime A}_a}\rangle$, 
leading to excitation of atom~$B$ from $|\Phi^B_{\gamma^B_b}\rangle$ to $|\Phi^B_{\gamma^{\prime B}_a}\rangle$
without change in the distribution of photon modes. 

\bigskip

\subsection{Single photons and sequentially/simultaneously emitted (absorbed) photon pairs}
\label{subsec:coupled-e-ph}
\subsubsection{One-photon exchange}
De-excitation of atom $B$ following the excitation triggers the dynamics 
of photon emission (absorption) dictated by the fourth term in Eq.~\eqref{eq:contributions}. The latter can be split according to the net number of exchanged photons as 
\begin{eqnarray}
	\label{eq:C1ph2ph}
	C^{(k+1)}_{[h\nu]}(\gamma^{\prime A}_a,\gamma^{\prime B}_b, \boldsymbol{n}^\prime;t)&=&
	-i\,\kappa_{1ph}\,\,C^{(k+1)}_{[\text{1ph}]}(\boldsymbol{\gamma}^\prime,\boldsymbol{n}^\prime;t)\nonumber\\[0.15cm] 
	&&-i\,\kappa_{2ph}\,\,C^{(k+1)}_{[\text{2ph}]}(\boldsymbol{\gamma}^\prime,\boldsymbol{n}^\prime;t),\quad
\end{eqnarray}
with $\kappa_{1ph}\!=\!-e/mc$ and $\kappa_{2ph}\!=\!e^2/2mc$. The first (second) term in Eq.~\eqref{eq:C1ph2ph} describes
the exchange of one (two) photons. The first term arises from the contribution  $\propto \Op{A}_s(\Op{r})\cdot\Op{p}$ in Eq.~\eqref{eq:H_I}.  It can be written as 
\begin{eqnarray}
\label{eq:C1ph}
	C^{(k+1)}_{[\text{1ph}]}(\boldsymbol{\gamma}^\prime,\boldsymbol{n}^\prime;t)&=& W^{(k+1)}_{[abs]}(\boldsymbol{\gamma}^\prime,\boldsymbol{n}^\prime;t)+W^{(k+1)}_{[em]}(\boldsymbol{\gamma}^\prime,\boldsymbol{n}^\prime;t).\!\,\,\nonumber\\ 
\end{eqnarray}
The two parts in Eq.~\eqref{eq:C1ph} dictate the absorption and emission of one net photon according to 
\begin{subequations}
\begin{widetext}
\begin{eqnarray}
\label{eq:W_abs}
	W^{(k+1)}_{[abs]}(\boldsymbol{\gamma}^\prime,\boldsymbol{n}^\prime ;t ) &=&
\!\sum_{\gamma^B_b}\sum_{q\in\Omega_\gamma}\,A_q\,\sqrt{n^\prime_q\!+\!1}\,\,
	\langle\Phi^B_{\gamma^{\prime B}_b}|e^{i\textbf{\text{k}}_q\cdot\Op{r}}\,\Op{p}|\Phi^B_{\gamma^{B}_b}\rangle\cdot\boldsymbol{\epsilon}_{q}
\int^{t}_{-\infty}\!\! S^{(k)}_{\!\gamma^{\prime A}_a,\gamma^{B}_b}(\boldsymbol{n}^\prime\!+\!\boldsymbol{1}_q ;t^\prime )\,e^{i(\epsilon^{B}_{\gamma^\prime_b}-\epsilon^{B}_{\gamma_b}-\hbar\omega_q\!)t^\prime}\,dt^\prime,
\end{eqnarray}
\vspace{-0.27cm}
\begin{eqnarray}
\label{eq:W_em}
	W^{(k+1)}_{[em]}(\boldsymbol{\gamma}^\prime,\boldsymbol{n}^\prime ;t ) &=&
\!\sum_{\gamma^B_b}\sum_{q\in\Omega_\gamma}\,A_q\,\sqrt{n^\prime_q}\,\,
	\langle\Phi^B_{\gamma^{\prime B}_b}|e^{-i\textbf{\text{k}}_q\cdot\Op{r}}\,\Op{p}|\Phi^B_{\gamma^{B}_b}\rangle\cdot\boldsymbol{\epsilon}_{q}
\int^{t}_{-\infty}\!\! S^{(k)}_{\!\gamma^{\prime A}_a,\gamma^{B}_b}(\boldsymbol{n}^\prime\!-\!\boldsymbol{1}_q ;t^\prime )\,e^{i(\epsilon^{B}_{\gamma^\prime_b}-\epsilon^{B}_{\gamma_b}+\hbar\omega_q\!)t^\prime}\,dt^\prime.\quad
\end{eqnarray}
\end{widetext}
\end{subequations}
Here we used the convention of Sec.~\ref{subsec:convention} and defined
$A_q=\sqrt{\hbar/2\omega_q\epsilon_0 V_0}$. The photon mode labelled as~$q$ is defined by 
the momentum, $\hbar\textbf{\text{k}}_q$,
and polarization, $\boldsymbol{\epsilon}_{q}$,  of the emitted (absorbed) photon.
Emission (absorption) of a photon in mode $q$ yields 
de-excitation (excitation) of atom $B$ from 
$|\Phi^B_{\gamma^{B}_b}\rangle$ to $|\Phi^B_{\gamma^{\prime B}_b}\rangle$.
For entangled photon pairs sequentially emitted in a radiative photon cascade process, the largest contributing term 
is given by Eq.~\eqref{eq:W_em}, although the dynamics of photoemission may also be affected by 
additional terms describing absorption and emission involving two-photon exchange processes.

\subsubsection{Two-photon exchange}
The coefficient $C^{(k+1)}_{[2ph]}(\boldsymbol{\gamma}^\prime,\boldsymbol{n}^\prime;t)$ in Eq.~\eqref{eq:C1ph2ph} arises 
from the term  proportional to $\Op{A}_s(\Op{r})\!\cdot\!\Op{A}_s(\Op{r})$ in Eq.~\eqref{eq:H_I}. It describes 
real and virtual two-photon emission (absorption) in the same or different modes with effective two or zero net photon exchange processes between atom~$B$ and the photon field. 
Compared to its counterpart proportional to $\kappa_{1ph}$ in Eq.~\eqref{eq:C1ph2ph}, this term is $\propto e^2$. Consequently, it provides a weaker contribution to one-step cascade processes. 
However, it must be taken into account for radiative processes involving two or more optical cascades, as the multiplying factors appearing in powers of~$e$ for both terms may become commensurate. 
We define this term according to whether or not the exchanged photons are in the same or different modes, namely
\begin{eqnarray}
	\label{eq:P+Q}
C^{(k+1)}_{[\text{2ph}]}(\boldsymbol{\gamma}^\prime,\boldsymbol{n}^\prime;t)&=& 
	P^{(k+1)}(\boldsymbol{\gamma}^\prime,\boldsymbol{n}^\prime;t) + Q^{(k+1)}(\boldsymbol{\gamma}^\prime,\boldsymbol{n}^\prime;t).\nonumber\\[0.0cm]
\end{eqnarray}
The first (second) term in Eq.~\eqref{eq:P+Q} describes the exchange of two photons in the same (or different) mode(s). The first term in Eq.~\eqref{eq:P+Q} may be decomposed as 
\begin{eqnarray}
P^{(k+1)}(\boldsymbol{\gamma}^\prime,\boldsymbol{n}^\prime;t)= P^{(k+1)}_{[em,em]}(\boldsymbol{\gamma}^\prime,\boldsymbol{n}^\prime;t)&+&P^{(k+1)}_{[abs,abs]}(\boldsymbol{\gamma}^\prime,\boldsymbol{n}^\prime;t)\nonumber\\
&+&P^{(k+1)}_{[em,abs]}(\boldsymbol{\gamma}^\prime,\boldsymbol{n}^\prime;t),\nonumber\\[0.0cm]
\end{eqnarray}
where we have defined (using $|\Phi^B_{\gamma^B_b}\rangle\equiv|\Phi^B_{\gamma_b}\rangle$),

\begin{subequations}
\begin{widetext}
\label{eq:Pall}
\begin{eqnarray}
\label{eq:P_abs_abs}
P^{(k+1)}_{[abs,abs]}(\boldsymbol{\gamma}^\prime,\boldsymbol{n}^\prime ;t ) &=&
\sum_{\gamma^B_b}\sum_{q\in\Omega_\gamma}\!A_q\,A_q\,   
\langle\Phi^B_{\gamma^\prime_b}|e^{ 2i\textbf{\text{k}}_q\cdot\Op{r}}|\Phi^B_{\gamma_b}\,\rangle\sqrt{n_q+1}\,\sqrt{n_q+2}\!
	\int^{t}_{-\infty}\!\!\!\!\! S^{(k)}_{\gamma^{\prime A}_a,\gamma^{B}_b}(\boldsymbol{n}^\prime\!+\!\boldsymbol{2}_q ;t^\prime )\, e^{i(\epsilon^{B}_{\gamma^\prime_b}-\epsilon^{B}_{\gamma_b}-2\hbar\omega_q )t^\prime}\!\!\!dt^\prime\,\,\quad\quad\,\,\,\,\,\\[-0.6cm]\nonumber
\end{eqnarray}
\begin{eqnarray}
\label{eq:P_em_em}
P^{(k+1)}_{[em,em]}(\boldsymbol{\gamma}^\prime,\boldsymbol{n}^\prime ;t ) &=&
\sum_{\gamma^B_b}\sum_{q\in\Omega_\gamma}\!A_q\,A_q\,
\langle\Phi^B_{\gamma^\prime_b}|e^{-2i\textbf{\text{k}}_q\cdot\Op{r}}|\Phi^B_{\gamma_b}\rangle\, \sqrt{n^\prime_q-1}\, \sqrt{n^\prime_q}
\int^{t}_{-\infty}\!\!\! 
	S^{(k)}_{\gamma^{\prime A}_a,\gamma^{B}_b}(\boldsymbol{n}^\prime\!-\!\boldsymbol{2}_q ;t^\prime )\,\, e^{i(\epsilon^{B}_{\gamma^\prime_b}-\epsilon^{B}_{\gamma_b}+2\hbar\omega_q )t^\prime}\! dt^\prime;\quad\quad\quad\\[-0.6cm]\nonumber
\end{eqnarray}
\begin{eqnarray}
\label{eq:P_abs_em}
P^{(k+1)}_{[em,abs]}(\boldsymbol{\gamma}^\prime, \boldsymbol{n}^\prime ;t ) &=&
\sum_{q\in\Omega_\gamma}\!A_q\,A_q\,(2n^\prime_q+1)\!\!\int^{t}_{-\infty}\!\!\!\! 
	S^{(k)}_{\gamma^{\prime A}_a,\gamma^{\prime B}_b}(\boldsymbol{n}^\prime;t^\prime )dt^\prime.\quad\quad\quad\quad\quad\quad\hspace{6.3cm}
\end{eqnarray}
\end{widetext}
\end{subequations}
Equation~\eqref{eq:P_abs_abs} describes the absorption of two photons in the same
mode~$q$ and subsequent excitation of atom~$B$ from $|\Phi^{B}_{\gamma^{B}_b}\rangle$ to $|\Phi^{B}_{\gamma^{\prime B}_b}\rangle$, leaving the scattered photoelectron unaffected
if two photons in the same mode~$q$ are present in the field at the previous iteration step $(k)$, and if the state $|\Phi^{B}_{\gamma^{B}_b}\rangle$ is populated. The distribution of modes then 
goes from $\boldsymbol{n}^\prime+\bold{2}_q$ to $\boldsymbol{n}^\prime_q$ photons in mode $q$, i.e., compare the terms $S^{(k)}_{\gamma^{\prime A}_a,\gamma^{B}_b}(\boldsymbol{n}^\prime+\boldsymbol{2}_q ;t^\prime )$ (uncorrected)
and $P^{(k+1)}_{[em,em]}(\boldsymbol{\gamma}^\prime,\boldsymbol{n}^\prime ;t)$ (corrected). 

Next, Eq.~\eqref{eq:P_em_em} accounts for the emission of two photons in the same 
mode~$q$. For $k_qr\ll1$,  the description of excitation (de-excitation) induced by simultaneous two-photon absorption (emission) requires
both terms to be treated beyond the dipole approximation. Note that emission and absorption of photons
may occur without excitation (de-excitation), even in the dipole approximation. Simultaneous emission of a photon in mode $q$ and absorption of a photon in the same mode  with no
electron dynamics involved is described by Eq.~\eqref{eq:P_abs_em}.

Finally, the second term in Eq.~\eqref{eq:P+Q}, describing the simultaneous exchange of two photons in different modes, reads 
\begin{eqnarray}
	Q^{(k+1)}(\boldsymbol{\gamma}^\prime,\boldsymbol{n}^\prime;t)\!=\! Q^{(k+1)}_{[abs,abs]}(\boldsymbol{\gamma}^\prime,\boldsymbol{n}^\prime;t)&\!+\!&Q^{(k+1)}_{[em,em]}(\boldsymbol{\gamma}^\prime,\boldsymbol{n}^\prime;t)\!\!\!\!\!\!\!\!\!\!\!\!\!\!\!\!\!\!\!\!\!\!\!\!\!\!\nonumber\\[0.1cm]
	&\!+\!&Q^{(k+1)}_{[em,abs]}(\boldsymbol{\gamma}^\prime,\boldsymbol{n}^\prime;t),\nonumber\\[0.0cm]
\end{eqnarray}
where, in analogy with Eq.~\eqref{eq:Pall},  we have defined
\begin{widetext}
\begin{subequations}
\label{eq:Qall}
\begin{eqnarray}
\label{eq:Q_abs_abs}
Q^{(k+1)}_{[abs,abs]}(\boldsymbol{\gamma}^\prime,\boldsymbol{n}^\prime ;t ) &=&
\sum_{\gamma^B_b}\sum_{\substack{p,q\in\Omega\\p\ne q}}
\!\!A^{p}_{q}(n^\prime_p\!+\!1,n^\prime_q\!+\!1)\,
\langle\Phi^B_{\gamma^\prime_b}|e^{i(\textbf{\text{k}}_p+\textbf{\text{k}}_q)\cdot\Op{r}} |\Phi^B_{\gamma_b}\rangle
\!\!\int^{t}_{-\infty}\!\!\!\!\! 
S^{(k)}_{\gamma^{\prime A}_a,\gamma^{B}_b}(\boldsymbol{n}^\prime\!\!+\!\!\boldsymbol{1}_p\!\!+\!\!\boldsymbol{1}_q;t^\prime ) 
	e^{i[\epsilon^{B}_{\gamma^\prime_b}-\epsilon^{B}_{\gamma_b}-\hbar(\omega_p+\omega_q)]t^\prime}\!dt^\prime; \nonumber\\[-0.6cm]
\end{eqnarray}
\begin{eqnarray}
\label{eq:Q_em_em}
Q^{(k+1)}_{[em,em]}(\boldsymbol{\gamma}^\prime,\boldsymbol{n}^\prime ;t ) &=&
\sum_{\gamma^B_b}\sum_{\substack{p,q\in\Omega\\p\ne q}}
\!\!A^{p}_{q}(n^\prime_p,n^\prime_q)\,
\langle\Phi^B_{\gamma^\prime_b}|e^{-i(\textbf{\text{k}}_p+\textbf{\text{k}}_q)\cdot\Op{r}} |\Phi^B_{\gamma_b}\rangle
\!\!\int^{t}_{-\infty}\!\!\!\!\! 
S^{(k)}_{\gamma^{\prime A}_a,\gamma^{B}_b}(\boldsymbol{n}^\prime\!+\!\boldsymbol{1}_p\!+\!\boldsymbol{1}_q;t^\prime )\,\, 
	e^{i[\epsilon^{B}_{\gamma^\prime_b}-\epsilon^{B}_{\gamma_b}+\hbar(\omega_p+\omega_q)]t^\prime}dt^\prime;\quad\quad\quad\\[-0.3cm]\nonumber
\end{eqnarray}
\begin{eqnarray}
\label{eq:Q_abs_em}
Q^{(k+1)}_{[abs,em]}(\boldsymbol{\gamma}^\prime,\boldsymbol{n}^\prime ;t ) &=&
\sum_{\gamma^B_b}\sum_{\substack{p,q\in\Omega\\p\ne q}}
\!\!A^{p}_{q}(n^\prime_p+1,n^\prime_q)\,
\langle\Phi^B_{\gamma^\prime_b}|e^{i(\textbf{\text{k}}_p-\textbf{\text{k}}_q)\cdot\Op{r}} |\Phi^B_{\gamma_b}\rangle
\!\!\int^{t}_{-\infty}\!\!\!\!\! 
S^{(k)}_{\gamma^{\prime A}_a,\gamma^{B}_b}(\boldsymbol{n}^\prime\!\!+\!\!\boldsymbol{1}_p\!\!+\!\!\boldsymbol{1}_q;t^\prime )\,\, 
e^{i[\epsilon^{B}_{\gamma^\prime_b}-\epsilon^{B}_{\gamma_b}+\hbar(\omega_q-\omega_p)]t^\prime}dt^\prime\,\,\,\,\,\quad\quad
\end{eqnarray}
\end{subequations}
\end{widetext}
with $A^{p}_{q}(n^\prime_p,n^\prime_q)= A_p\,A_q\, \sqrt{n^\prime_p}\,\sqrt{n^\prime_{q}}\,\,(\textbf{\text{e}}_p\cdot\textbf{\text{e}}_q)$. 
Equation~\eqref{eq:Q_abs_abs} accounts for the simultaneous absorption of two photons in different modes, while
Eq.~\eqref{eq:Q_em_em} describes the simultaneous emission of two photons in different modes. Finally, Eq.~\eqref{eq:Q_abs_em}
accounts for the exchange of photon pairs without a net change in the photon number: it describes 
the simultaneous absorption and emission of one photon in different modes. The matrix elements and  angular distributions 
involving simultaneously emitted (absorbed) photon pairs are evaluated following
the prescription given in Sec.~\ref{subsubsec:ADaTME2}.

For the photon wavelengths considered in this work, the contribution 
due to the term $\Op{A}_s(\Op{r})\cdot\Op{A}_s(\Op{r})$ dri\-ving \textit{simultaneous photon pair emission}
via optical de-excitation is expected to be less important compared to the case of \textit{sequential photon pair 
emission} mediated by the term $\Op{A}_s(\Op{r})\cdot\Op{p}$,  
since the matrix elements corresponding to the former would vanish in the dipole approximation.
It is worth mentioning that Eq.~\eqref{eq:W_em}, or more specifically, 
a two-step sequential application of Eq.~\eqref{eq:W_em}, provides the leading contribution 
to the emission of photon pairs with energies $h\nu_1$ and $h\nu_2$: the correlated photon pair
originates from sequential emission from the target atom~$B$, 
following the de-excitation pathways depicted in Fig.~1(c) of the manuscript.
\subsection{Exchange term $C^{(k+1)}_{[X]}(\gamma^{\prime A}_a,\gamma^{\prime B}_b,\boldsymbol{n}^\prime;t)$}
\label{subsection:exchange_term}

Finally, proper antisymmetrization of the two-electron wave function is ensured by the last term 
in Eq.~\eqref{eq:contributions}: $C^{(k+1)}_{[X]}(\gamma^{\prime A}_a,\gamma^{\prime B}_b,\boldsymbol{n}^\prime;t)$.
The latter updates the expansion coeffients to ensure the required exchange symmetry via the iterative correction 
\begin{subequations}
\begin{widetext}
\begin{eqnarray}
\label{eq:CX}
C^{(k+1)}_{[X]}(\gamma^{\prime A}_a,\gamma^{\prime B}_b,\boldsymbol{n}^\prime;t)&=&
	i\sum_{\gamma^{A}_a,\gamma^{B}_b}\langle\Phi^{A}_{\gamma^{\prime A}_a}|\Op{V}_{ne}(\Op{r}\!-\!R_{B})|\Phi^B_{\gamma^B_b}\rangle\langle\Phi^{B}_{\gamma^{\prime B}_b}|\Phi^{A}_{\gamma^{A}_a}\rangle
	\int^{t}_{-\infty}\!\!S^{(k)}_{\gamma^{A}_a,\gamma^{B}_b}(\boldsymbol{n}^\prime ;t^\prime ) e^{i(\epsilon^{A}_{\gamma^\prime_a}+\epsilon^{B}_{\gamma^\prime_b}-\epsilon^A_{\gamma_a}-\epsilon^{B}_{\gamma_b}) t^\prime}\, dt^\prime\nonumber\\[0.1cm]
&+&i\sum_{\gamma^{A}_a,\gamma^{B}_b}\langle\Phi^{A}_{\gamma^{\prime A}_a}\Phi^{A}_{\gamma^{\prime B}_b} |\Op{V}_{\tiny\!ee}(\Op{r},\Op{r}^\prime)|\Phi^B_{\gamma^B_b}\Phi^A_{\gamma^A_a}\rangle
\int^{t}_{-\infty}\!\!S^{(k)}_{\gamma^{A}_a,\gamma^{B}_b}(\boldsymbol{n}^\prime ;t^\prime ) e^{i(\epsilon^{A}_{\gamma^\prime_a}+\epsilon^{B}_{\gamma^\prime_b}-\epsilon^A_{\gamma_a}-\epsilon^{B}_{\gamma_b}) t^\prime}\, dt^\prime\nonumber\\[0.1cm]
	&-&\,\,\sum_{\gamma^A_a}\,\,\underline{\underline{C}}^{(k+1)}_{\,[h\nu]}(\gamma^A_a,\gamma^{\prime B}_{b},\boldsymbol{n}^\prime;t)
\,+\!\sum_{\gamma^A_a,\gamma^B_b}
	\mathcal{O}^{\gamma^{\prime A}_a,\gamma^{\prime B}_b}_{\gamma^{A}_a,\gamma^{B}_b}
\left[S^{(k)}_{\gamma^{A}_a,\gamma^{B}_b}(\boldsymbol{n}^\prime;t^\prime)\,e^{(\epsilon^{A}_{\gamma^\prime_a}+\epsilon^{B}_{\gamma^\prime_b}-\epsilon^{A}_{\gamma_a}-\epsilon^{B}_{\gamma_b})t^\prime}\right]^{t}_{-\infty}\!\!,\quad\quad
\end{eqnarray}
\end{widetext}
where the overlap matrix, already appearing in Eq.~\eqref{eq:S1}, reads
\begin{eqnarray}	
	\mathcal{O}^{\gamma^{\prime A}_a,\gamma^{\prime B}_b}_{\gamma^{A}_a,\gamma^{B}_b}&=&
\langle\Phi^A_{\gamma^{\prime A}_a}|\Phi^B_{\gamma^{B}_b}\rangle\langle\Phi^B_{\gamma^{\prime B}_b}|\Phi^A_{\gamma^A_a}\rangle. 
\end{eqnarray}	
\end{subequations}
The double underline in the third term in Eq.~\eqref{eq:CX}
indicates that the expressions 
$(\epsilon^{B}_{\gamma^{\prime}_b}-\epsilon^{B}_{\gamma_b})$
as well as  all transition matrix elements of the form  
$\langle\Phi^{B}_{\gamma^{\prime B}_b}|\mathcal{T}|\Phi^{B}_{\gamma^{B}_b}\rangle$
appearing though Eqs.~\eqref{eq:W_abs}$-$\eqref{eq:Q_abs_em}
must be replaced with $(\epsilon^{A}_{\gamma^{\prime}_{a}}+\epsilon^{B}_{\gamma^{\prime}_{b}}-\epsilon^{A}_{\gamma_{a}}+\epsilon^{B}_{\gamma_{b}})$
and $\langle\Phi^{B}_{\gamma^{\prime B}_b}|\mathcal{T}|\Phi^{A}_{\gamma^{A}_a}\rangle\langle\Phi^{A}_{\gamma^{\prime A}_{a}}|\Phi^{A}_{\gamma^{B}_{b}}\rangle$, res\-pectively.
Note that the classical field does not contribute to the exchange term 
because of $f_{\Omega_A}(r)$ and the localized character of $\mathcal{H}_B$, i.e., 
the expansion coefficients vanish for the continuum states of atom~$B$:
ionization of atom $B$ by electron impact is not considered. The maximum number of sequential radiative cascade steps describing the de-excitation pathways from an excited state to the ground state  
is determined by the maximum order
in the perturbation expansion, $k_{max}$. The reported results were obtained by
iterating the expansion coefficients up to $k_{max}\!=\!6$, corresponding to the lowest order needed
to describe the process of radiative two-photon cascade emission while considering 
feedback effects of the emitted photons on the scattered electron wave packet.

\section{Polarization states and propagation direction of emitted (absorbed) photons}
\label{sec:transversality}
\subsection{Transversality condition and polarization components}
\subsubsection{General scheme}

Throughout the text, we used the shorthand notation 
\begin{eqnarray}
\label{eq:myconv}
\boldsymbol{\sigma}_j\equiv(\sigma_{ja},\sigma_{jb}),\quad
\end{eqnarray}
when referring to the indices of the two mutually orthogonal polarization components $\hat{\epsilon}_{\sigma_{ja}}$ and $\hat{\epsilon}_{\sigma_{jb}}$ satisfying the transversality
conditions
\begin{eqnarray}
\label{eq:transversality}
\hat{\epsilon}_{\boldsymbol{\sigma}_j}\cdot\textbf{\text{k}}_j=0.\quad 
\end{eqnarray}
Within this convention, summation over the indices $\boldsymbol{\sigma}_j$, e.g., in Eq.~\eqref{eq:Psi_S_k}, implies summation over
their components in Eq.~\eqref{eq:myconv} for each momentum $\hbar\textbf{\text{k}}_j$. 
To construct $\epsilon_{\boldsymbol{\sigma}_{j}}$, or equivalently  $\epsilon_{\sigma_{ja}}$ and $\epsilon_{\sigma_{jb}}$ 
satisfying Eq.~\eqref{eq:transversality}, we write $\textbf{\text{k}}_j$ in polar coordinates,
\begin{eqnarray}
\label{eq:kpolar}
	\textbf{\text{k}}_j &=& \sqrt{\dfrac{4\pi}{3}}\,\left(\dfrac{\omega_j}{c}\right)\!\sum_{q=0,\pm 1}\,\text{Y}^{1}_{q}(\theta_{_{k_j}},\phi_{_{k_j}} )\,\hat{\textbf{\text{e}}}^{*}_q, 
\end{eqnarray}
with $Y^1_q(\theta_{k_j},\phi_{k_j})$ as the spherical harmonic for the propa\-gation direction of the mode $(\textbf{\text{k}}_j,\boldsymbol{\sigma}_j)$ 
and $\hat{\textbf{\text{e}}}_q$ denoting the covariant spherical unit vectors 
$\mathbf{e}_{-1}\!=\!(\mathbf{e}_{x}\!-\!i\mathbf{e}_y)/\sqrt{2}$, 
$\mathbf{e}_{+1}\!=\!-(\mathbf{e}_{x}\!+\!i\mathbf{e}_y)/\sqrt{2}$, and $\mathbf{e}_{0}\!=\!\mathbf{e}_z$.  We then construct 
the orthogonal triad of unit vectors $(\textbf{\text{k}}_j/|\textbf{\text{k}}_j|,\epsilon_{\sigma_{jq}},\epsilon_{\sigma_{jq}})$ as 

\begin{eqnarray}
	\hat{\epsilon}_{\sigma_{ja}}=\dfrac{\textbf{\text{k}}_j\times\hat{\textbf{\text{e}}}_0}{|\textbf{\text{k}}_j\!\times\!\hat{\textbf{\text{e}}}_0|}
	&=&\dfrac{1}{\sqrt{2}} \Big[i\cos(\theta_{k_j}) - \sin(\phi_{k_j})\Big]\,\textbf{\text{e}}_{+1}\nonumber\\
	&+&
	\dfrac{1}{\sqrt{2}}\Big[i\cos(\theta_{k_j}) + \sin(\phi_{k_j})\Big]\,\textbf{\text{e}}_{-1},~~~
\end{eqnarray}
for the first polarization component and
\begin{eqnarray}
	\hat{\epsilon}_{\sigma_{jb}}&=&\dfrac{\textbf{\text{k}}_j\times\hat{\epsilon}_{\sigma_{ja}}}{|\textbf{\text{k}}_j\!\times\!\hat{\epsilon}_{\sigma_{ja}}|}\nonumber\\
	&=&-\dfrac{1}{\sqrt{2}} \Big[i\sin(\phi_{k_j})\cos(\theta_{k_j}) +\cos(\phi_{k_j})\cos(\theta_{k_j})\Big]\,\textbf{\text{e}}_{+1}\nonumber\\
	&&-\dfrac{1}{\sqrt{2}} \Big[i\sin(\phi_{k_j})\cos(\theta_{k_j}) -\cos(\phi_{k_j})\cos(\theta_{k_j})\Big]\,\textbf{\text{e}}_{-1}\nonumber\\
	&&-\sin(\theta_{k_j})\,\textbf{\text{e}}_{0},
\end{eqnarray}
for the second polarization component, with
$(\times)$ denoting the vector cross product. They both fulfill the transversality condition $\hat{\boldsymbol{\epsilon}}_{\boldsymbol{\sigma}_j}\cdot\textbf{\text{k}}_j=0$.   

In  cartesian coordinates, 
$[\hat{\boldsymbol{e}}_{x},\hat{\boldsymbol{e}}_{y},\hat{\boldsymbol{e}}_{y}]$,
$\epsilon_{\sigma_{ja}}$ and $\epsilon_{\sigma_{jb}}$ are 
\begin{eqnarray}
\label{eq:pol_cartesian}
	\hat{\epsilon}_{\sigma_{ja}}\!\equiv\!\begin{bmatrix}
	\sin(\phi_{k_j})\\[0.2cm]
	-\cos(\phi_{k_j})\\[0.2cm]
0
\end{bmatrix};	
\hspace{0.2cm}
	\hat{\epsilon}_{\sigma_{jb}}\!\equiv\!    
\begin{bmatrix}
\cos(\phi_{k_j})\cos(\theta_{k_j}) \\[0.2cm]
-\sin(\phi_{k_j})\cos(\theta_{k_j})\\[0.2cm]
 -\sin(\theta_{k_j})
\end{bmatrix}\!\!.\quad
\end{eqnarray}

\subsubsection{Application to the correlated photodetection scheme}
\label{subsubsec:application_corr_scheme}
Following the photodetection scheme of Fig.~(2b) in the manuscript, a photon in mode (2) 
is detected at the fixed angles $\theta_{k_2}=\pi/2$ and $\phi_{k_2}=-\pi/2$. This gives, according to Eq.~\eqref{eq:pol_cartesian}, 
$\epsilon_{\sigma_{2a}}=[-1,0,0]$ and $\epsilon_{\sigma_{2b}}=[0,0,-1]$. The first detector 
measures the photon with energy $h\nu_2$ with polarization component along $\epsilon_{\sigma_{2b}}$, i.e., along $\hat{z}$.  

A second detector detects the photon with energy $h\nu_1$ at fixed $\phi_{k_1}=\pi/2$, but it is free to move and hence scan
along $\theta_{k_1}$. According to Eq.~\eqref{eq:pol_cartesian}, this corresponds to 
$\epsilon_{\sigma_{1a}}=[1,0,0]$ and $\epsilon_{\sigma_{1b}}=[0,-\cos(\theta_{k_1}),-\sin(\theta_{k_1})]$. 
The second detector 
measures the correlated photon pair with energy $h\nu_1$
with polarization component along $\epsilon_{\sigma_{1b}}$ as it scans along $\theta_{k_1}$.\\

\subsection{Angular distributions and transition matrix elements}
\label{subsec:ADaTME}
\subsubsection{Angular distributions of single photons and sequentially emitted (absorbed) photon pairs}
\label{subsubsec:ADaTME1}
The directions of propagation of the emitted (absorbed) photons 
are evaluated by means of a multipole expansion of the exponential terms, $\exp(ikr)$, 
which appear in the matrix elements of the expansion coefficients defined throughout Sec.~\ref{subsec:coupled-e-ph}.  Specifically,
\begin{eqnarray}
e^{\pm i\textbf{\text{k}}_{j}\cdot\textbf{\text{r}}}\!=\!4\pi\!\sum_{\lambda,\mu}\!(\pm i)^{\lambda} j_\lambda(k_j r)Y^{\lambda}_{\mu}(\theta_r,\phi_r)\, Y^{\lambda*}_{\mu}(\theta_{k_j},\phi_{k_j}\!),\nonumber \\
\end{eqnarray}
where $j_\lambda(k_j r)$ denotes a spherical Bessel function.
For the photon energies considered in this work, the wavelengths defining the modes emitted and
absorbed during the propagation of the expansion coefficients are several orders of magnitude 
larger than the extension of the most diffuse bound state of the target atom $B$.
The Bessel functions can therefore be approximated by 
\begin{equation}
j_\lambda(k_jr)\approx(k_j r)^{\lambda}/(2\lambda+1)!!, 
\end{equation}
since $k_j r \ll 1$.  This allows us to obtain the transition matrix elements as a power series in $r^\lambda$.
Following this prescription, the matrix elements defined in Eqs.~\eqref{eq:W_abs} and~\eqref{eq:W_em} 
are given by
\begin{subequations}
\begin{eqnarray}
	\langle\Phi^{B}_{\gamma^{\prime B}_b} | e^{\pm\textbf{\text{k}}_j\cdot\Op{r}}  \Op{p}_{\mu_0}|\Phi^B_{\gamma^{B}_b}\rangle &\!=\!&
	\sum_{\lambda,\mu} 
	\langle\Phi^{B}_{\gamma^{\prime B}_b} | \mathcal{D}^{[\lambda,\pm]}_{\mu,\mu_0}(k_j)|\Phi^B_{\gamma^{B}_b}\rangle\nonumber\\
	&&\times\text{Y}^{\lambda *}_{\mu}(\Omega_{k_j}),
\end{eqnarray}
with $\Omega_{k_j}\!\equiv\!(\theta_{k_j},\phi_{k_j})$. The multipole coefficients are obtained by standard angular-momentum techniques as
\begin{widetext}
\begin{eqnarray}
	\small
	\langle\Phi^{B}_{\gamma^{\prime B}_b} | \mathcal{D}^{[\lambda,\pm]}_{\mu,\mu_0}(k_j)|\Phi^B_{\gamma^{B}_b}\rangle&=&
	C\, \dfrac{(\pm i)^{\lambda}\,(k_j)^\lambda}{(2\lambda+1)!!} (-1)^{l^\prime+m^\prime} 
	\!\!\!\!\!\sum^{l^\prime+\lambda}_{q=|l^\prime-\lambda|}\!\left[\dfrac{(2l^\prime+1)(2\lambda+1)}{1/(2q+1)}\right]^{\frac{1}{2}}\!
	\begin{pmatrix}
		l^\prime&\lambda&q\\[0.10cm]
	 -m^\prime&\mu&m+\mu_0
	\end{pmatrix}
	\begin{pmatrix}
		l^\prime&\lambda&q\\[0.10cm]
	 0&0&0
	\end{pmatrix}
	\delta_{m^\prime-\mu,m+\mu_0}\nonumber\\[0.3cm]
	&\times&
	\small\left\{ 
         \begin{matrix}	
	 &\!\!\!\!\!\!
	\displaystyle\int^{+\infty}_{0} 
	\!\!\!r^{\lambda+2} 
	f_{n^\prime,l^\prime}(r)\left(\dfrac{\partial}{\partial r} - \dfrac{l}{r}\right) f_{n,l}(r) dr \left[\left(\dfrac{l+1}{2l+3}\right)^{\!\frac{1}{2}}
        \!(-1)^{l-1+m+\mu_0}	
	\begin{pmatrix}
		l&1&l+1\\[0.15cm]
	 m&\mu_0&-m-\mu_0
	\end{pmatrix}
	\right]\!\delta_{q,l+1}\\[0.78cm]
	-&\!\!
	\displaystyle\int^{+\infty}_{0} 
	\!\!\!r^{\lambda+2} f_{n^\prime,l^\prime}(r)\left(\dfrac{\partial}{\partial r} + \dfrac{l}{r}\right) f_{n,l}(r) dr\! \left[\left(\dfrac{l}{2l-1}\right)^{\!\frac{1}{2}}
        \!\!(-1)^{l-1+m+\mu_0}	
	\begin{pmatrix}
		l&1&l-1\\[0.15cm]
	 m&\mu_0&\,\,\,\,\,m+\mu_0
	\end{pmatrix}
	\right]\!\delta_{q,l-1}
	 \end{matrix}\right.\nonumber	 
		\\ 
\end{eqnarray}
\end{widetext}
\end{subequations}
with $C=-i\hbar\sqrt{36\pi}$.  Furthermore, $(n^\prime,l^\prime,m^\prime)$ and $(n,l,m)$
denote the quantum numbers corresponding to $\gamma^{\prime B}_b$ and $\gamma^{B}_b$, 
while $f_{n^\prime,l^\prime}(r)$ and $f_{n,l}(r)$ are the radial components of the orbital wavefunctions 
$\langle\boldsymbol{r}|\Phi^{B}_{\gamma^{\prime B}_{b}}\rangle$ and $\langle\boldsymbol{r}|\Phi^{B}_{\gamma^{B}_{b}}\rangle$, respectively.

\subsubsection{Angular distributions of simultaneously emitted (absorbed) photon pairs}
\label{subsubsec:ADaTME2}

Analogously, the matrix elements describing simultaneous two-photon exchange, 
i.e., emission ($s\!=\!-1$) or absorption ($s\!=\!+1$)
of a photon in mode $q_s$ with momentum $\hbar\textbf{\text{k}}_s$ and  emission ($s^\prime=-1$) or absorption ($s^\prime=+1$) of another photon in mode $q_{s^\prime}$ with momentum $\hbar\textbf{\text{k}}_{s^\prime}$, 
such as those 
appearing in Eqs.~\eqref{eq:Qall}, are obtained according to the multipole expansion involving the product of two spherical harmonics.  Specifically,
\begin{subequations}
\begin{eqnarray}
	\langle\Phi^{B}_{\gamma^{\prime B}_b} | e^{i(s\textbf{\text{k}}_s + {s^\prime}\textbf{\text{k}}_{s^\prime})\cdot\Op{r}} |\Phi^B_{\gamma^{B}_b}\rangle &\!=\!&
	\!\!\!\sum_{\substack{ \lambda_{s^\prime},\,\mu_{s^\prime}\\\lambda_s,\,\mu_s}}\!\!\! 
	\langle\Phi^{B}_{\gamma^{\prime B}_b} | \mathcal{O}^{[\lambda_s,\lambda_{s^\prime}]}_{\mu_s,\mu_{s^\prime}}(\boldsymbol{v}^{s^\prime}_{s})|\Phi^B_{\gamma^{B}_b}\rangle\nonumber\\[0.2cm]
	&\times& \text{Y}^{\lambda_s *}_{\mu_s}(\Omega_{k_s})\, \text{Y}^{\lambda_{s^\prime} *}_{\mu_{s^\prime}}(\Omega_{k_{s^\prime}})\,,\quad
\end{eqnarray}
	where we have defined $\boldsymbol{v}^{s^\prime}_{s}\equiv[k_s,s,k_s^\prime,s^\prime]$. The coefficients are again obtained after straightforward angular-momentum algebra.  The result is
\begin{widetext}
\begin{eqnarray}
	\small
	\langle\Phi^{B}_{\gamma^{\prime B}_b} | \mathcal{O}^{[\lambda_s,\lambda_{s^\prime}]}_{\mu_s,\mu_{s^\prime}}(\boldsymbol{v}^{s^\prime}_{s})|\Phi^B_{\gamma^{B}_b}\rangle&=&
	 (4\pi)^2\dfrac{(is)^{\lambda_s}\,(is^\prime)^{\lambda_{s^\prime}}}{(2\lambda_s+1)!!}	
	 \dfrac{(k_s k_{s^{\prime}})^{\lambda_s+\lambda_{s^\prime}}}{(2\lambda_{s^\prime}+1)!!}\,
	(-1)^{l^\prime+m^\prime-\mu_s-\mu_{s^\prime}} 
	 \displaystyle\int^{+\infty}_{0}\!\!\! f_{n^\prime\!,l^\prime}(r)\,\,r^{\lambda_{s}+\lambda_{s^\prime}+2}\,\,f_{n,l}(r)\, dr\nonumber\\[0.2cm]
	 &\times&\sum_{\Lambda=|\lambda_s-\lambda_{s^\prime}|}^{\lambda_s+\lambda_{s^\prime}}\!\!\!(-1)^{\Lambda} 
	\begin{pmatrix}
        \lambda_s&\lambda_{s^\prime}&\Lambda\\[0.15cm]
        \mu_s&\mu_{s^\prime}&-\mu_s-\mu_{s^\prime}
	\end{pmatrix}
	\begin{pmatrix}
        \lambda_s&\lambda_{s^\prime}&\Lambda\\[0.15cm]
        0&0&0
	\end{pmatrix}
	\begin{pmatrix}
        l^\prime&\Lambda&l\\[0.15cm]
		-m^\prime&\mu_s+\mu_{s^\prime}&m
	\end{pmatrix}
	\begin{pmatrix}
        l^\prime&\Lambda&l\\[0.15cm]
		0&0&0
	\end{pmatrix}\nonumber\\[0.2cm]
	&\times&
\left[\dfrac{(2\lambda_s+1)(2\lambda_{s^\prime}+1)}{4\pi/(2\Lambda+1)}\right]^{\frac{1}{2}}
	\delta_{m^\prime-m,\mu_s+\mu_{s^\prime}}.
\end{eqnarray}
\end{widetext}
\end{subequations}

%


\end{document}